\documentclass[reprint,aip,amssymb,amsmath,cha]{revtex4-1}

\usepackage{amsfonts,amssymb,amsmath}
\usepackage{amsthm, enumerate, graphicx, url}

\usepackage{bm}%
\usepackage[colorlinks=true,linkcolor=blue]{hyperref}%
\expandafter\ifx\csname package@font\endcsname\relax\else
 \expandafter\expandafter
 \expandafter\usepackage
 \expandafter\expandafter
 \expandafter{\csname package@font\endcsname}%
\fi
\def\R{\mathbb{R}}
\def\E{\mathbb{E}}
\def\Z{\mathbb{Z}}
\def \tblb{\\ \hline}

\begin{document}

\title{A finite difference method for estimating second order parameter sensitivities of discrete stochastic chemical reaction networks}

\author{Elizabeth Skubak Wolf}
\email{skubak@math.wisc.edu}
\affiliation{Department of Mathematics, University of Wisconsin, Madison, Wisconsin 53706, USA}

\author{David F. Anderson}
\email{anderson@math.wisc.edu}
\affiliation{Department of Mathematics, University of Wisconsin, Madison, Wisconsin 53706, USA}

\begin{abstract}
  We present an efficient finite difference method for the approximation of 
second derivatives, with
respect to  system parameters, of  expectations  for a class of discrete stochastic chemical reaction networks. The method uses a coupling of the perturbed processes that yields a much lower variance than existing methods, thereby drastically lowering
the computational complexity required to solve a given problem. Further, the method is simple to implement and will also prove useful in any setting in which continuous time Markov chains are used to model dynamics, such as population processes.  We expect the new method to be useful in the context of optimization algorithms that require knowledge of the Hessian.
\end{abstract}

\maketitle
\section{Introduction}

Stochastic models are commonly used to simulate and analyze chemical and biochemical systems, in particular when the abundances of the constituent molecules are small and ordinary differential equations cease to provide a good description of  system behavior.  The most common modeling choice is to use a continuous time Markov chain (CTMC), which we represent via the stochastic equation \eqref{model_notheta}, but is often described in the biology literature via the chemical master equation, and simulated using Gillespie's algorithm\cite{Gill76,Gill77} or the next reaction method.\cite{Gibson2000,Anderson2007a}

Parameter sensitivity analysis is a valuable tool in this setting as it provides a quantitative method for understanding how perturbations in the parameters affect different response functions of interest.  Further, often the only means of determining the parameters for these models is experimentally.  If the model provides a reasonable approximation of the system, the sensitivities can be used to analyze the identifiability of given parameters. \cite{Komorowski2011}  They can also suggest an experimental design in which more resources, which are often limited, can be spent determining the more sensitive parameters.

While first derivative sensitivities have been much studied, less focus has been given to finding reasonable algorithms for the computation of sensitivities of higher order, particularly in the discrete state setting.  Second derivative sensitivities (the Hessian), however, are also useful.  For example, they provide concavity information which is necessary for finding roots or extrema of an expectation.  Additionally, in a more general optimization setting, the Hessian can be used to improve upon a simple steepest-descent method.  Newton and quasi-Newton methods, for instance, use an approximate Hessian to choose the direction in which to step in the next iterate of the optimization, using curvature to find a more direct path to a local minimum than can be achieved by using the gradient alone.  When the Hessian is positive semi-definite, these methods achieve a fast rate of local convergence.  Additionally, trust-region based optimization methods can also be markedly improved by including a Hessian estimate.\cite{GlynnAsmussen2007,Nocedal2006,Rawlings2002,Spall2009}
Developing algorithms that successfully integrate these optimization methods in the chemical reaction network and CTMC setting, for example in the context of parameter estimation, is a topic of current research which depends critically on having an efficient method for approximating second derivatives.

We introduce here a method for the computation of the second order sensitivities of stochastically modeled biochemical reaction networks that is a nontrivial extension of the coupled finite difference method developed in \onlinecite{Anderson2011}.  The proposed method produces an estimate with a significantly lower variance than existing methods, so that it requires much less CPU time to produce an approximation within a desired tolerance level. 
Additionally, the paths generated can also be re-used to compute first derivatives of the system for use in any optimization algorithm.
While biochemical reaction networks will be the setting for this paper, the proposed method is also applicable to a wide variety of continuous time Markov chain models, such as those used in queueing theory and the study of population processes.

The outline of the paper is as follows.  In Section \ref{sec:model}, we precisely describe the model and problem under consideration.   In Section \ref{method}, we present the new method and give a simple algorithm for implementation.  In Section \ref{numexamples}, we provide several numerical examples to compare the new method with existing methods, including finite differencing with common random numbers, finite differencing with the common reaction path method, and second order likelihood transformations. Finally, in Section \ref{sec:conclusion}, we provide some conclusions and discuss avenues for future work.

\section{The Formal Model}
\label{sec:model}

Suppose we have a system of $d$ chemical species undergoing $M$ reactions, each with a given propensity function $\lambda_k:\R^d\to \R_{\ge0}$ (known as an intensity function in the mathematical literature) and reaction vector (transition direction) $\zeta_k\in\mathbb{R}^d$.    We can model this system as a continuous time Markov chain using the random time change representation  \cite{Kurtz72,Kurtz86,AndKurtz2011}
\begin{equation}\label{model_notheta}
X_t=X_0 +\sum_{k=1}^M Y_k\left(\int_0^t
\lambda_k(X_s)ds\right)\zeta_k,
\end{equation}
where the $Y_k$ are independent, unit-rate Poisson processes and $X_0$ is the initial state. 
We assume, without loss of generality, that the state space $\mathcal{S}$ is some subset of $\mathbb{Z}^d_{\geq 0}$.  That is, the abundances of the constituent species are always non-negative integers.  Note that the chemical master equation (forward equation in the language of probability) for the above model is
\begin{align*}
	\frac{d}{dt}P_{X_0}(x,t) &= \sum_{k=1}^M P_{X_0}(x - \zeta_k,t)\lambda_k(x-\zeta_k)1_{\{x-\zeta_k \in \Z^d_{\ge 0}\}} \\
	&\hspace{.1in}- P_{X_0}(x,t)\sum_{k=1}^M \lambda_k(x),
\end{align*}
where $P_{X_0}(x,t)$ is the probability of being in state $x\in \Z^d_{\ge 0}$ at time $t\ge 0$ given an initial condition of $X_0.$

  Intuitively, the random time change representation \eqref{model_notheta} can be understood as follows.
Let $R_k(t):=Y_k\left(\int_0^t \lambda_k(X_s)ds\right)$.  
Then $R_k(t)$ counts the number of times the $k$th reaction has occurred up to time $t$, and $R_k(t)\zeta_k$ is the change in the system due to these reactions.  The representation (\ref{model_notheta}) then shows that the process at time $t$ is simply its initial value plus the total change up to time $t$ due to each of the $M$ reactions.
To understand the counting processes $R_k(t)=Y_k\left(\int_0^t \lambda_k(X_s)ds\right)$, picture a realization of the unit-rate Poisson process $Y_k$ as being determined by points on $\R_{\geq 0}$ giving the jump times of $Y_k$.\footnote{The gaps between points are unit exponential random variables.}  For example, we could have
\begin{center}\
  \begin{tabular}{|l|l} \hspace{0.5in}x \hspace{0.1in}x
    \hspace{0.8in}x \hspace{0.4in}x& \hspace{0.2in}x
    \hspace{0.3in}
    \tblb
    &$s$ \\
  \end{tabular}\ 
\end{center}
\noindent where the ``X'' marks correspond to the jump times of the Poisson process.  
At time zero begin at the origin, and as time increases, travel to the right. At a given time $s$ on the line, the value $Y_k(s)$ is how many points we have passed up to and including $s$.  For example, in the picture above, $Y_k(s) = 4$.  The propensity function $\lambda_k$ then indicates how fast we travel on this line.  If $\lambda_k(X_s)$ is very large and $t_1>t_0$, we expect $\int_0^{t_1} \lambda_k(X_s)ds$ to be much larger than $\int_0^{t_0} \lambda_k(X_s)ds$, so that $Y_k\left(\int_0^{t_1} \lambda_k(X_s)ds\right)$ is much larger than $Y_k\left(\int_0^{t_0} \lambda_k(X_s)ds\right)$; we have traveled past many points along the line between times $t_0$ and $t_1$, and so were ``moving quickly.''  At the other extreme, if $\lambda_k(X_s)$ is zero between $t_0$ and $t_1$, then $\int_0^{t_1} \lambda_k(X_s)ds=\int_0^{t_0} \lambda_k(X_s)ds$, so that $Y_k\left(\int_0^{t_1} \lambda_k(X_s)ds\right)=Y_k\left(\int_0^{t_0} \lambda_k(X_s)ds\right)$; in this case we did not travel anywhere on the line, but were ``stopped.''
For further information, intuition, and a derivation of this representation, see  \onlinecite{Kurtz82, AndKurtz2011, Kurtz86}.

Now we suppose that the propensities are dependent on some vector of parameters $\theta$; for instance, $\theta$ may represent a subset of the system's mass action kinetics constants.  We then consider a family of  models $X_t(\theta)$, parameterized by $\theta$, with stochastic equations 
\begin{equation}\label{model}
X_t(\theta)=X_0(\theta)+\sum_{k=1}^M Y_k\left(\int_0^t
\lambda_k(\theta,X_s(\theta))ds\right)\zeta_k.
\end{equation}
Letting $f$ be some function of interest, for example the abundance of some molecule, we define
\[
	J(\theta) := \E f(\theta,X_{t}(\theta)).
\]
This paper is concerned with finding an efficient computational method for the approximation of the second partial derivatives of $J$,
\[
	\frac{\partial^2}{\partial\theta_j\partial\theta_i}J(\theta).
\]

There are several existing methods for the approximation of such sensitivities.  The likelihood ratio (LR) method proceeds analytically by moving the derivative inside the expectation. \cite{Plyasunov2007,GlynnAsmussen2007}  The variance of such estimators, however, are often prohibitive.  In Section \ref{numexamples}, we include numerical results from the LR method for comparison.  The general method of infinitesimal perturbation (IP) also proceeds by moving the derivative inside the expectation. \cite{Khammash2012}  However, IP methods do not apply for many stochastically modeled chemical reaction networks, as the requirements of the needed analytical tools are typically not met.   See the appendix of \onlinecite{Khammash2012}.

Finite difference methods for approximating these sensitivities start with the simple observation that for smooth functions $J$, 
we may approximate second partial derivatives by perturbing the parameter vector in both relevant directions, so that
\begin{align}
\label{basicdiff}
	&\frac{\partial^2}{\partial\theta_j\partial\theta_i}J(\theta)  \\
	&= \frac{J(\theta+(e_i+e_j)\epsilon) -J(\theta+e_i\epsilon) - J(\theta+e_j\epsilon)+ J(\theta)}{\epsilon^2} \notag \\
	&\hspace{.2in} + O(\epsilon) \notag,
\end{align}
where $e_i$ is the vector with a 1 in the $i^{th}$ position and 0 elsewhere.  Thus, for second derivatives, finite difference methods require up to four simulated paths to produce one estimate, as opposed to the LR method, which requires only one path per estimate.  When coupling methods are used with the finite difference, however, the variance of the estimates produced are usually significantly lower than LR, as demonstrated in Section \ref{numexamples}, so that finite different methods often provide much more effective estimators.

In our setting, equation \eqref{basicdiff} suggests an approximation of the form
\begin{widetext}
\begin{align}
\frac{\partial^2}{\partial\theta_j\partial\theta_i}J(\theta)& \approx \E \left( \frac{f(\theta,X_t(\theta+(e_i+e_j)\epsilon))-f(\theta,X_t(\theta+e_i\epsilon))-f(\theta,X_t(\theta+e_j\epsilon))+f(\theta,X_t(\theta))}{\epsilon^2} \right)\label{diff} .
\end{align}
\end{widetext}
The Monte Carlo estimator for  (\ref{diff}) with $R$ estimates is then
\begin{equation}\label{eq:estimator}
	D_R(\epsilon)=\frac{1}{R}\sum_{\ell=1}^R d_{[\ell]}(\epsilon),
\end{equation}
where 
\begin{align*}
 &d_{[\ell]}(\epsilon) \\
 &\hspace{.1in}:=\epsilon^{-2}\big[ f(\theta,X_{t,[\ell]}(\theta+(e_i+e_j)\epsilon)) -f(\theta,X_{t,[\ell]}(\theta+e_i\epsilon))\\
 &\hspace{.3in}-f(\theta,X_{t,[\ell]}(\theta+e_j\epsilon))+f(\theta,X_{t,[\ell]}(\theta))\big],
\end{align*}
where, for example, $X_{t,[\ell]}(\theta)$ is the $\ell$th path simulated with parameter choice $\theta$. Note that if the four relevant processes are computed independently, which we will call the Independent Random Numbers (IRN) method, the variance of the estimator $D_R(\epsilon)$ is $R^{-1}\textsf{Var}(d(\epsilon)) = O(R^{-1}\epsilon^{-4})$.  The goal of any coupling method in this context is to lower the variance of $d(\epsilon)$ by correlating the relevant processes.  

We will demonstrate via example that the method presented here lowers the variance of the numerator of $d(\epsilon)$ to $O(\epsilon)$, thereby lowering the variance of $d(\epsilon)$ to $O(\epsilon^{-3})$, yielding  $\textsf{Var}(D_R(\epsilon)) = O(R^{-1}\epsilon^{-3})$.   The proof of this fact follows from  work in \onlinecite{Anderson2011}.  For several non-trivial examples, however, the method gives even better performance, lowering the variance of $d(\epsilon)$ another order of magnitude to $O(\epsilon^{-2})$.
In contrast, every other coupling method we attempted\footnote{We do not provide a full list of the less efficient couplings, of which there are many.} yielded an asymptotic variance for $d(\epsilon)$ of $O(\epsilon^{-3})$ at best, and in general were much less efficient than the method being proposed here. 
Theoretical work, and a discussion of conditions on the model guaranteeing the faster rate of convergence, will be presented in a follow-up paper.

For ease of exposition and notation, we have described finite differences using the forward difference \eqref{diff}.  Our formal construction will also use the forward difference.  In practice, however, it is no more difficult to use the central second difference,
\begin{align} \label{cdiff}\begin{split}
&\epsilon^{-2}\big[ f(\theta,X_t(\theta+(e_i+e_j)\epsilon/2))-f(\theta,X_t(\theta+(e_i-e_j)\epsilon/2))\\
& -f(\theta,X_t(\theta+(e_j-e_i)\epsilon/2))+f(\theta,X_t(\theta-(e_i+e_j)\epsilon/2))\big],
\end{split}
\end{align}
which has a bias of only $O(\epsilon^2)$; this is what we have implemented in our numerical examples.

\section{Coupling the Finite Difference}
\label{method}

The goal of any coupling of the finite difference is to reduce the variance of the estimator produced by somehow ensuring that the four paths needed in \eqref{basicdiff}  remain close together.  The common random numbers (CRN) coupling achieves this goal by reusing the uniform random numbers in an implementation of Gillespie's direct algorithm.\cite{Gill77}  Implicit in equation \eqref{model} is the Common Reaction Path (CRP) coupling, \cite{Khammash2010} which assigns a stream of random numbers to each $Y_k$ which are then used to produce the required realizations of the stochastic processes.

The method presented here, on the other hand, forces the paths to share reactions; often two or more of the four paths have the same reaction occur at the same point in time.  Further, the method often naturally ``recouples'' the processes during the course of a simulation.\cite{Anderson2011} These facts allow the paths to remain closer than is possible by only sharing random numbers, and so the method consistently produces an estimate with lower variance, often significantly so.  
We provide numerical evidence for this comparison of methods in Section \ref{numexamples}; we also briefly revisit this discussion at the end of Section \ref{construction}.

\subsection{Review of first derivatives}\label{sec:review}
The main idea of the method presented here is most easily seen in the context of first derivatives, where only correlated pairs of runs are required to approximate the first finite difference  
\[
	\epsilon^{-1} (J(\theta + e_i \epsilon)-J(\theta)).
\]
The main idea of the coupling presented in this paper is illustrated in the following toy example.  Suppose we wish to study the difference between two Poisson processes $Z_1,Z_2$ with rates 13.1 and 13, respectively.  One way would be to use independent, unit rate Poisson processes $Y_1, Y_2$ and write
\begin{equation}\label{toy}
 Z_1(t)=Y_1(13.1t) \ \ \textrm{and} \ \  Z_2(t)=Y_2(13t).
 \end{equation}
 Then $\mathbb{E}(Z_1(t)-Z_2(t))=0.1t$ and
$\textrm{Var}(Z_1(t)-Z_2(t))=\textrm{Var}(Y_1(13.1t)) + \textrm{Var}(Y_2(13t))=26.1t.$

We would like to lower this variance: instead, write
\[ Z_1(t)=Y_1(13t)+Y_2(0.1t) \ \ \textrm{and} \ \  Z_2(t)=Y_1(13t).\]
Then we still have
$\mathbb{E}(Z_1(t)-Z_2(t))=\mathbb{E}Y_2(0.1t)=0.1t$ as needed, but now
$\textrm{Var}(Z_1(t)-Z_2(t))=\textrm{Var}(Y_2(0.1t))=0.1t$ instead.

In general, we want to consider the difference of two processes $Z_1$, $Z_2$ with intensities $A$ and $B$ (the intensities may be functions of time, but this is not important to the main idea).  In this case, we would write
\[ Z_1= Y_1 + Y_2 \ \ \textrm{and} \ \  Z_2 = Y_1 + Y_3,\]
where $Y_1, Y_2,$ and $Y_3$ are independent unit-rate Poisson processes that have intensities 
\[
	m:=\min\{A,B\}, \ A-m,  \text{ and } B-m,
	\]
	 respectively.  In other words, we have split the counting processes $Z_1$ and $Z_2$ into three sub-processes.  One of these, $Y_1$, is shared between $Z_1$ and $Z_2$, so that at the time at which $Y_1$ jumps, both of the original processes jump. These shared jumps lower the variance of the difference $Z_1-Z_2$ to only $A+B-2m = |A-B|$, rather than the $A+B$ which would result from using only two processes similarly to \eqref{toy} above.

This is precisely the idea needed for the computation of first derivatives, as in \onlinecite{Anderson2011}, which is discussed in Section \ref{alt} in more detail.  This idea will also serve as the basis for our coupling method for second derivatives.

\subsection{Construction of the Coupling}\label{construction}

For the computation of second derivites, rather than correlated pairs of runs, correlated quartets of runs are used.
We suppose we have the four CTMCs of \eqref{diff}, with $i,j\in\{1,\dots,M\}$ fixed, which for convenience of exposition we order as 
\begin{equation}\label{order}
X_t(\theta+(e_i+e_j)\epsilon), \ X_t(\theta+e_i\epsilon),\
X_t(\theta+e_j\epsilon), \ X_t(\theta).
\end{equation}
 We also assume that their initial conditions are equal (i.e., they are equal at $t=0$), to some value $X_0(\theta)$.  For each of the four processes above, there is an associated propensity for each of the $M$ reaction channels.  For example, the propensity of the $k$th reaction channel of the first process (the one with parameter choice $\theta + (e_i + e_j)\epsilon$) is
\[	
	 \lambda_{k,1}:= \lambda_k(\theta + (e_i + e_j)\epsilon, X_t(\theta + (e_i + e_j)\epsilon)).
\]Similarly rename the propensity of the $k$th reaction channel of the second process (parameter choice $\theta + e_i\epsilon$) by $\lambda_{k,2}$, the third process as $\lambda_{k,3}$, and the fourth process as $\lambda_{k,4}$, as per our ordering (\ref{order}).  Note that these propensities are dependent on $\theta$ and $X_t(\theta)$, but we will drop either or both of these dependencies in our notation when they are not relevant to the current discussion.

Next, we introduce a coupling of these four processes that will produce an estimator \eqref{eq:estimator} with low variance.  The main idea is similar to that in Section \ref{sec:review} as well as in \onlinecite{Anderson2011} in that it rests on splitting a  counting process  into sub-processes, to be shared among the four CTMCs \eqref{order}.
With this goal in mind, we create a sub-process to allow the 1st and 2nd processes of \eqref{order} to jump simultaneously, one to allow the 1st and 3rd to jump simultaneously, one for the 2nd and 4th, and one for the 3rd and 4th.  Additionally, we create a sub-process that allows all four to jump simultaneously.  As in the first derivative setting, the rates of these sub-processes will involve minimums of the original CTMCs.  Finally, we also require four additional sub-processes to make up any ``leftover'' propensity of the original CTMCs.

Formally, define $R_{k,[b_1,b_2,b_3,b_4]}$ as a counting process, where $b_\ell \in \{0,1\}$.  A jump of $R_{k,[b_1,b_2,b_3,b_4]}$ indicates that the $\ell$th process in the ordering (\ref{order}) jumps by reaction $k$ if and only if $b_\ell=1$, for $\ell\in\{1,2,3,4\}$. For example, $R_{k,[1,1,0,0]}(t)$ counts the number of times the $k$th reaction has fired simultaneously for the first and second processes of \eqref{order} (but the third and fourth did not fire), whereas $R_{k,[1,0,1,0]}(t)$ counts the number of times the $k$th reaction has fired simultaneously for the first and third processes of \eqref{order} (but the second and fourth did not fire).  Define the propensity of $R_{k,[b_1,b_2,b_3,b_4]}$ by $\Lambda_{k,[b_1,b_2,b_3,b_4]}$, so that in the random time change representation (\ref{model}),
\begin{equation}\label{rdefn}
 R_{k,[b_1,b_2,b_3,b_4]}(t) = Y_{k,[b_1,b_2,b_3,b_4]}\left( \int_0^t \Lambda_{k,[b_1,b_2,b_3,b_4]}(s) ds \right)
 \end{equation}
where the $Y$'s are independent unit-rate Poisson processes and where the propensities are
\begin{align}
&
\Lambda_{k,[1,1,1,1]} = \lambda_{k,1}\wedge\lambda_{k,2}\wedge\lambda_{k,3}\wedge\lambda_{k,4} \notag \\ \notag
&
\Lambda_{k,[1,1,0,0]} = \lambda_{k,1}\wedge\lambda_{k,2} - \Lambda_{k,[1,1,1,1]} \\ \notag
&
\Lambda_{k,[0,0,1,1]} = \lambda_{k,3}\wedge\lambda_{k,4} - \Lambda_{k,[1,1,1,1]} \\ \notag
&
\Lambda_{k,[1,0,1,0]} = (\lambda_{k,1}- \lambda_{k,1}\wedge\lambda_{k,2} )\wedge (\lambda_{k,3} -  \lambda_{k,3}\wedge\lambda_{k,4})\\ \label{intensities}
&
\Lambda_{k,[0,1,0,1]} = (\lambda_{k,2}- \lambda_{k,1}\wedge\lambda_{k,2} )\wedge (\lambda_{k,4} -  \lambda_{k,3}\wedge\lambda_{k,4})\\ \notag
&
\Lambda_{k,[1,0,0,0]} = (\lambda_{k,1}-\lambda_{k,1}\wedge\lambda_{k,2}) - \Lambda_{k,[1,0,1,0]}\\ \notag
&
\Lambda_{k,[0,1,0,0]} = (\lambda_{k,2}- \lambda_{k,1}\wedge\lambda_{k,2} ) - \Lambda_{k,[0,1,0,1]}\\ \notag
&
\Lambda_{k,[0,0,1,0]} = (\lambda_{k,3} -  \lambda_{k,3}\wedge\lambda_{k,4}) - \Lambda_{k,[1,0,1,0]}\\ 
&
\Lambda_{k,[0,0,0,1]} = (\lambda_{k,4} -  \lambda_{k,3}\wedge\lambda_{k,4})-\Lambda_{k,[0,1,0,1]} \notag,
\end{align}
where we define the notation $a\wedge b := \min\{a,b\}$.
The proposed coupling is then given by the following:
\begin{widetext}
\begin{align}
\begin{split}
X_t(\theta+(e_i+e_j)\epsilon) &= X_0(\theta)+ \sum_k \zeta_k 
(R_{k,[1,1,1,1]}(t) + R_{k,[1,1,0,0]}(t) + R_{k,[1,0,1,0]}(t) + R_{k,[1,0,0,0]}(t) )
\\
X_t(\theta+e_i\epsilon) &=X_0(\theta)+ \sum_k \zeta_k 
(R_{k,[1,1,1,1]}(t) + R_{k,[1,1,0,0]}(t) + R_{k,[0,1,0,1]}(t) + R_{k,[0,1,0,0]}(t) )
\\
X_t(\theta+e_j\epsilon) &= X_0(\theta)+ \sum_k \zeta_k
(R_{k,[1,1,1,1]}(t) + R_{k,[0,0,1,1]}(t) + R_{k,[1,0,1,0]}(t) + R_{k,[0,0,1,0]}(t) )
\\
X_t(\theta) &= X_0(\theta)+ \sum_k \zeta_k
(R_{k,[1,1,1,1]}(t) + R_{k,[0,0,1,1]}(t) + R_{k,[0,1,0,1]}(t) + R_{k,[0,0,0,1]}(t) ). 
\end{split}
\label{coupling}
\end{align}
\end{widetext}
A few comments are in order.  First, note that, for example, the marginal process $X_t(\theta+(e_i+e_j)\epsilon)$  above  involves all the counting processes in which $b_1=1$.  Second, each of these marginal processes $X_t(\cdot)$ have the same distribution as the original, uncoupled, processes since the transition rates of the marginal processes have remained unchanged.  This can be checked by simply summing the rates of the relevant counting processes, which are all
 those $\Lambda_{k,[b_1,b_2,b_3,b_4]}$ in which a given $b_\ell = 1$.
Third,  if $f$ is linear, for example if we are estimating the abundance of a particular molecule, many of the $R_{k,[b_1,b_2,b_3,b_4]}$ are completely cancelled if we now construct the difference (\ref{diff}).  An example of this will be shown in Section \ref{simpleex}.
Fourth,  even if $i=j$, the coupling requires two different copies of the process $X_t(\theta + e_i\epsilon)$, one taking the role of $X_t(\theta+e_i\epsilon,t)$ and the other $X_t(\theta+e_j\epsilon)$. 


As discussed at the beginning of this section, the CRN and CRP methods attempt to reduce the variance of the estimator \eqref{eq:estimator} by  reusing random numbers for each of the four nominal processes.
However, as discussed in \onlinecite{Anderson2011} in the setting of first derivatives, this will often lead to a decoupling over long enough time periods. Hence, the variance of the CRN and CRP estimators will often eventually converge to a variance of the same order of magnitude as the estimator constructed using independent samples.  This behavior is demonstrated by example in the current setting of second derivatives in Section \ref{numexamples}.  The double coupled method presented here 
\textit{re-couples} the four relevant processes every time they are near each other, which, by contrast, does not  occur in either CRN or CRP.
We refer the interested reader to \onlinecite{Anderson2011}, Section 3.1 for a more thorough discussion of this idea.

\subsection{An Alternative Derivation}\label{alt}
The coupling described in the previous section can be derived in an alternate way, which explains why the method is termed ``double coupled.''  
We could first couple the first and second processes of \eqref{order} using the coupled finite difference method,  \cite{Anderson2011} and then couple the third and fourth in the same manner.    For example, using the $\lambda_{k,\ell}$ as defined in the previous section, the first two processes in (\ref{order}) are constructed as:
\begin{small}
\begin{align}
\begin{split}
X(\theta +(e_i+e_j)\epsilon,t) &= X_0(\theta) +  \sum_k \left( R_{k,[1,1]} + R_{k,[1,0]} \right) \zeta_k \\
X(\theta+e_i\epsilon,t) &= X_0(\theta)+ \sum_k \left( R_{k,[1,1]} + R_{k,[0,1]} \right)\zeta_k,
\end{split}
\label{eq:first_diff}
\end{align}
\end{small}
where $R_{k,[b_1,b_2]}= Y_{k,[b_1,b_2]}\left(\int_0^t \Lambda_{k,[b_1,b_2]}(s) ds\right)$ are defined analogously to (\ref{rdefn}) and where
\begin{align*}
 \Lambda_{k,[1,1]}(s) &= \lambda_{k,1}(s)\wedge\lambda_{k,2}(s),\\
\Lambda_{k,[1,0]}(s) &= \lambda_{k,1}(s) - \lambda_{k,1}(s)\wedge\lambda_{k,2}(s),\\
\Lambda_{k,[0,1]}(s) &= \lambda_{k,2}(s) - \lambda_{k,1}(s)\wedge\lambda_{k,2}(s).
\end{align*}

As in \eqref{coupling}, the processes defined in \eqref{eq:first_diff} jump together as often as possible: they share the sub-processes $R_{k,[1,1]}$, each of which runs at a propensity equal to the minimum of the respective propensities of the two original processes.  We then expect the variance of the first finite difference $[f(\theta,X_t(\theta+(e_i+e_j)\epsilon)) -f(\theta,X_t(\theta+e_i\epsilon))]\epsilon^{-1}$ to be small since  the two processes of \eqref{eq:first_diff} will remain approximately the same whenever they jump simultaneously via $R_{k,[1,1]}$.

Now note that, together, the two processes \eqref{eq:first_diff} can be viewed as a new 
CTMC with dimension $2d$, twice that of that of the original process.  
The third and fourth processes in (\ref{order}) can be similarly coupled, giving us \textit{two} $2d$-dimensional CTMCs.  Finally, we couple these new processes into a single CTMC of dimension $4d$, in precisely the same manner of \onlinecite{Anderson2011}.  This construction leads to the same process as given in (\ref{coupling}).  The details are left to the interested reader.

\subsection{Algorithms for simulation of \eqref{coupling}}\label{algorithm}
We present two algorithms for the pathwise simulation of the equations \eqref{coupling}.  The first corresponds to the next reaction method of \onlinecite{Anderson2007a}, whereas the second corresponds to an implementation of Gillespie's direct method. \cite{Gill76,Gill77}   As usual, it will be problem specific as to which algorithm is most efficient.

Below, rand(0,1) indicates a uniform[0,1] random variable, independent from all previous random variables.  Recall that if $U\sim$ rand(0,1), then $\ln(1/U)/\lambda$ is exponentially distributed with parameter $\lambda >0$.  Also recall that even if $i$ and $j$ are equal, the processes $X(\theta+e_i\epsilon)$ and $X(\theta+e_j\epsilon)$ are still constructed separately.  Define the set
 \begin{align*}
 	B := &\{ [1,1,1,1], [1,1,0,0], [0,0,1,1], [1,0,1,0], \\
	&[0,1,0,1], [1,0,0,0], [0,1,0,0], [0,0,1,0], [0,0,0,1] \} 
\end{align*}
	and note that it will often be convenient to use a \textit{for} loop, from 1 to 9, to enumerate over the vectors in $B$.

\bigskip

\noindent\textsc{Algorithm-modified next reaction method applied to \eqref{coupling}}. \\
\textbf{Initialization}: Set $X(\theta+(e_i+e_j)\epsilon)=X(\theta+e_i\epsilon)=X(\theta+e_j\epsilon)
=X(\theta)=X_0$ and $t=0$; for each $k\in\{1,\dots,M\}$ and each $b\in B$, set $T_{k,b}=0$ and  $P_{k,b}=\ln(1/u_{k,b})$ for $u_{k,b}\sim$ rand(0,1).
\bigskip

\noindent \textbf{Repeat} the following steps:
\begin{enumerate}[(i)]
\item For each $k$, set 
\begin{align*}
\lambda_{k,1} &= \lambda_k(\theta + (e_i + e_j)\epsilon, X(\theta + (e_i + e_j)\epsilon)) \\
\lambda_{k,2} &= \lambda_k(\theta + e_i\epsilon, X(\theta + e_i\epsilon))
\\
 \lambda_{k,3} &=\lambda_k(\theta + e_j\epsilon, X(\theta + e_j\epsilon))
\\
 \lambda_{k,4} &= \lambda_k(\theta, X(\theta))
\end{align*}
and use to set each of the nine variables $\Lambda_{k,b}$ as above in (\ref{intensities}).
\item For each $k$ and $b\in B$, set 
\[
	\Delta t_{k,b} = \left\{ \begin{array}{ccc}
	 (P_{k,b}-T_{k,b})/\Lambda_{k,b} & , & \text{if } \Lambda_{k,b} > 0\\
	 \infty & , & \text{else }
	 \end{array}\right. .
\]
\item Set $\Delta = \min_{k,b}\{\Delta t_{k,b}\}$ and let $\mu:=k$ and $\nu:=b=[b_1,b_2,b_3,b_4]$ be the indices where the minimum is achieved.
\item Set $t=t+\Delta$.
\item Update state vector variables $X(\theta+(e_i+e_j)\epsilon),X(\theta+e_i\epsilon),X(\theta+e_j\epsilon),X(\theta) $ by adding $\zeta_\mu$ to the $\ell$th process if and only if $b_\ell=1$ in $\nu$.
\item For each $k$ and $b\in B$, set $T_{k,b}=T_{k,b}+\Delta \cdot \Lambda_{k,b}.$
\item Set $P_{\mu,\nu}=P_{\mu,\nu}+\ln(1/u)$ where $u\sim$ rand(0,1).
\item Return to $(i)$ or quit.
\end{enumerate}

\bigskip

\noindent\textsc{Algorithm-Gillespie's Direct Method Applied to \eqref{coupling}}. \\
\textbf{Initialization}: Set $X(\theta+(e_i+e_j)\epsilon)=X(\theta+e_i\epsilon)=X(\theta+e_j\epsilon)
=X(\theta)=X_0$ and $t=0$.
\bigskip

\noindent \textbf{Repeat} the following steps:
\begin{enumerate}[(i)]
\item For each $k$, set 
\begin{align*}
\lambda_{k,1} &= \lambda_k(\theta + (e_i + e_j)\epsilon, X(\theta + (e_i + e_j)\epsilon)) \\
\lambda_{k,2} &= \lambda_k(\theta + e_i\epsilon, X(\theta + e_i\epsilon))
\\
 \lambda_{k,3} &=\lambda_k(\theta + e_j\epsilon, X(\theta + e_j\epsilon))
\\
 \lambda_{k,4} &= \lambda_k(\theta, X(\theta))
\end{align*}
and use to set each of the nine variables $\Lambda_{k,b}$ as above in (\ref{intensities}).
\item Let $\Lambda_0 = \sum_{k}\sum_{b} \Lambda_{k,b}$ and $u \sim \text{rand}(0,1)$, and set
\[
	\Delta = \ln(1/u)/\Lambda_0.
\]
\item Set $t=t+\Delta$.
\item Let $u\sim \text{rand}(0,1)$ and use to select $(\mu,\nu)\in \{(k,b):k\in \{1,\dots,M\}, b\in B\}$ where each pair $(k,b)$ is selected with probability $\lambda_{k,b}/\Lambda_0$.\footnote{This is the usual generation of a discrete random variable found in every instance of Gillespie's algorithm.}

\item Update state vector variables $X(\theta+(e_i+e_j)\epsilon),X(\theta+e_i\epsilon),X(\theta+e_j\epsilon),X(\theta) $ by adding $\zeta_\mu$ to the $\ell$th process if and only if $b_\ell=1$ in $\nu$.
\item Return to $(i)$ or quit.
\end{enumerate}

\section{Numerical Examples}\label{numexamples}
In this section, we compare the double coupled method with the following existing methods:
\begin{enumerate}[(a)]

\item the usual Independent Random Numbers (IRN) estimator in which the processes of \eqref{diff} are simulated independently, also referred to as the crude Monte Carlo method,

\item the common random numbers approach (CRN) in which the processes of \eqref{diff} are simulated given the same stream of random numbers using Gillespie's direct algorithm,\footnote{At each step, the first random number determines the time of the next reaction, and the second determines which occurs; the reactions were listed in a fixed order as given in this paper.}

\item the Common Reaction Path (CRP) method of \onlinecite{Khammash2010} in which the processes of \eqref{diff} are coupled by  reusing each $Y_k$ of \eqref{model},

\item the double coupled method proposed here (CFD2, where the CFD stands for ``coupled finite difference'') which implements the coupling (\ref{coupling}),

\item a Girsanov transformation or likelihood ratio method (LR) in which the computed weight function is used as a control variate (see \onlinecite{Plyasunov2007}, and \onlinecite{GlynnAsmussen2007} Chapters V.2, VII.3).

\end{enumerate}  
All methods except (b) were simulated using the next reaction algorithm, modified as necessary.  
 We also note that the first four methods use the second finite difference, which has some bias (see Section \ref{sec:model}); recall that to reduce this bias we actually simulate the centered difference (\ref{cdiff}), which is accomplished in the same way as the forward difference but with the parameters shifted.    The LR method is the only one of the four methods we use here that is unbiased; its high variance, however, typically makes the method unusable.  Finally, when discussing performance, we will refer to $R$ of \eqref{eq:estimator} as the number of estimates.

\subsection{A Simple Birth Process}\label{simpleex}

Consider a pure birth process $A \to 2A.$
Here, $\zeta = 1$, and denoting by $X_t$ the number of $A$ molecules at time $t$, we assume a propensity function $\lambda(\theta,X_t(\theta))$, so that
in the random time change representation,
\[X_t(\theta) = X_0 + Y\left(\int_0^t \lambda(\theta,X_s(\theta))ds\right),\]
where, as usual, $Y$ is a unit-rate Poisson process. 

Suppose we are interested in the second derivative of $\E X_t$ with respect to $\theta$ (so that $f(\theta,x)=x$). We double couple the processes as in (\ref{coupling}), noting that we are in the special case when $i=j$. This does not change the main idea of the double coupling, but it requires us to distinguish the two nominal processes with the same parameter value $\theta + \epsilon$; we label them as $X_t^1(\theta+\epsilon)$ and $X_t^2(\theta+\epsilon)$.  Ordering as in (\ref{order}), and noting that since there is only one reaction we may drop the subscript $k$, we find that 
\begin{align*}
\lambda_{1}&=\lambda(\theta+2\epsilon,X_t(\theta+2\epsilon))\\
\lambda_{2}&=\lambda(\theta+\epsilon,X_t^1(\theta+\epsilon))\\
\lambda_{3}&=\lambda(\theta+\epsilon,X_t^2(\theta+\epsilon))\\
\lambda_{4}&=\lambda(\theta,X_t(\theta)),
\end{align*}
and use these to define the $\Lambda$'s as given in (\ref{intensities}).  The double coupled processes are then given as
\begin{widetext}
\begin{align*}
X_t(\theta+2\epsilon) &= X_0(\theta)+ 
R_{[1,1,1,1]}(t) + R_{[1,1,0,0]}(t) + R_{[1,0,1,0]}(t) + R_{[1,0,0,0]}(t) 
\\
X_t^1(\theta+\epsilon) &=X_0(\theta)+
R_{[1,1,1,1]}(t) + R_{[1,1,0,0]}(t) + R_{[0,1,0,1]}(t) + R_{[0,1,0,0]}(t) 
\\
X_t^2(\theta+\epsilon) &= X_0(\theta)+ R_{[1,1,1,1]}(t) + R_{[0,0,1,1]}(t) + R_{[1,0,1,0]}(t) + R_{[0,0,1,0]}(t) 
\\
X_t(\theta) &= X_0(\theta)+ R_{[1,1,1,1]}(t) + R_{[0,0,1,1]}(t) + R_{[0,1,0,1]}(t) + R_{[0,0,0,1]}(t) .
\end{align*}
\end{widetext}
Now that we have coupled the processes, note that when we consider the second difference (\ref{diff}) for the given $f$, which is linear, most of the sub-processes cancel.  For example, since $R_{[1,1,0,0]}$ is present in both $X_t(\theta+2\epsilon)$, which is positive in the difference, and in $X_t^1(\theta+\epsilon)$, which is negative, $R_{[1,1,0,0]}$ is not present in the second difference.  One can easily check that the numerator of the difference \eqref{diff} simplifies in this case to
\begin{align}\label{eq:another}
\begin{split}
	&X_t(\theta+2\epsilon)-X^1_t(\theta+\epsilon)-X^2_t(\theta+\epsilon)+X_t(\theta) \\
	&= R_{[1,0,0,0]}(t) -R_{[0,1,0,0]}(t) - R_{[0,0,1,0]}(t) + R_{[0,0,0,1]}(t) .
	\end{split}
\end{align}
Note that  the rates of the four remaining counting processes of \eqref{eq:another} are usually relatively small; in fact, at any given time at least two of the four must have zero propensity, as can be seen by considering the possible values of the minima involved.



Suppose that $\lambda(\theta,X_t(\theta))=\theta X_t(\theta)$ is simply a constant times the population at time $t$.  We choose to estimate $\frac{\partial^2 \E X_t}{\partial\theta^2}$ at $t=5$ and  $\theta= 1/2$, with $X_0(\theta)= 1$.   We use $\epsilon= 1/50$ for the finite difference methods.  For simple examples such as this, one can solve for the derivative explicitly; in this case the actual value is 304.6.

As can be seen in the data in Table \ref{table:simplelin}, manifested in the width of the confidence interval, the variance of the double coupled estimator is  smaller than that of the estimators given by the other methods. 
For instance, for the same number of estimates it gives a confidence interval of half the width of the CRP and CRN methods, which for this single-reaction model, though implemented differently, give equivalent estimators.  Here and throughout, confidence intervals are constructed as $\pm1.96\sqrt{v}$ where $v$ is the variance of the estimator  \eqref{eq:estimator}.  

\begin{table}
\centering 
\begin{tabular}{|c|c|c|c|c|} 
\hline 
Method & Estimates & Approximation & \# updates & CPU time (s) \\ [0.5ex] 
\hline\hline
IRN & 100,000 & 307 $\pm$ 447 & $\approx 3.7 \times 10^6 $ &  38 \\ 
\hline
CRP & 100,000 & 315 $\pm$ 24 & $\approx 3.7 \times 10^6 $& 49 \\ 
\hline
CRN & 100,000 & 282 $\pm$ 24 & $\approx 3.3 \times 10^6 $& 32 \\ 
\hline
LR & 100,000 & 311 $\pm$ 20 &$\approx 1.1 \times 10^6 $ &  37 \\ 
\hline
CFD2 & 100,000 & 296 $\pm$ 12 &$\approx 1.2 \times 10^6 $ & 22  \\ 
\hline
\end{tabular}

\caption{95\% confidence intervals and computation time for each of the five methods (a) through (e), after 100,000 estimates, on the simple birth model of \ref{simpleex} (with linear propensity). An $\epsilon$ of 1/50 was used for the three finite difference methods.  Actual value: 304.6.}
\label{table:simplelin} 
\end{table}

For each method, we also include the CPU time that was required for the simulation, as well as the number of updates made to the system state (the number of times a reaction vector is added to the state vector).  The latter is a useful comparison tool, as it provides a measure of the amount of work the method requires, but is not influenced by  differences in implementation (such as use of Gillespie vs next reaction algorithms).  These differences, on the other hand, often affect CPU time. We do not also provide a random number count for each method, but note here that except for CRN this number is equal to the number of system updates.  For CRN, which uses Gillespie's algorithm, two random numbers are used per system update.  Finally, the CPU time will certainly vary by machine; all tests described in this section were run in MATLAB on a Windows machine with a 1.6GHz processor.

\subsection{mRNA Transcription and Translation }

We now examine the performance of the proposed method on a more realistic model.  In the following model of gene transcription and translation, mRNA is being created, and then translated into protein, while both the mRNA and the protein may undergo degradation (where here the constants are in the sense of mass action kinetics, so for example protein is being created at a rate of $\gamma$ times the number of mRNA molecules):
$$\emptyset \underset{\theta}{\overset{2}{\rightleftarrows}} M \hspace{.15cm} {\overset{\gamma}{\rightarrow}} \hspace{.15cm}M + P, \hspace{.5cm} P\hspace{.15cm}{\overset{1}{\rightarrow}} \hspace{.15cm} \emptyset.$$

We assume initial concentrations of zero mRNA and protein molecules.  The stochastic equation for this model is 
\begin{widetext}
\begin{align*}
X(\theta,t)= \ &Y_1(2t)\left( \begin{smallmatrix} 1\\ 0 \end{smallmatrix} \right)+Y_2\left(\int_0^t \theta X_M(\theta,s)ds\right)\left( \begin{smallmatrix} -1\\  0 \end{smallmatrix} \right)  +Y_3\left(\int_0^t \gamma X_M(\theta,s)ds\right)\left( \begin{smallmatrix} 0\\ 1 \end{smallmatrix} \right) + Y_4\left(\int_0^t X_P(\theta,s)ds\right)\left( \begin{smallmatrix} 0\\ -1 \end{smallmatrix} \right)
\end{align*} 
\end{widetext}
where $X = \left( \begin{smallmatrix} X_M\\ X_P \end{smallmatrix} \right)$ gives the numbers of the mRNA and protein molecules respectively.  Note that we have moved the parameter $t$ from the subscript for notational convenience.

In subsection \ref{mrna}, we compute the second derivative of the expected number of protein molecules with respect to $\theta$, while in subsection \ref{mixed}, we compute the mixed partial of this same quantity with respect to both $\theta$ and $\gamma$.  In subsection \ref{genf}, we compute the second derivative of the \textit{square} of the expected number of protein molecules with respect to $\theta$.

\subsubsection{$2^{nd}$ derivative of protein abundance with respect to $\theta$}\label{mrna}

Suppose we would like to estimate the second derivative of the expected number of protein molecules with respect to $\theta$ at a time of $t=30$ and $\theta= \frac{1}{4}$.  Additionally, we fix $\gamma=10$ and $X_0=0$.  One can analytically find that $\frac{\partial^2}{\partial\theta^2}\mathbb{E}X_P(30) = 2496$.

First, Table \ref{table:mp1} gives simulation data as in the previous examples, with two different perturbations, $\epsilon$, of $\theta$ used.  Note the trade-off between bias and precision: a larger epsilon implies the second finite difference has a larger bias, but, since there is an $\epsilon^2$ in the denominator of the estimator, the variance of the estimator is smaller; for small epsilon it is vice-versa.  Table \ref{table:mp2} shows the relevant data for the LR method.  

\begin{table*}
\centering 
\begin{tabular}{|c|c|c|c|c|c|} 
\hline 
Method & Estimates & $\epsilon = 1/20$ & $\epsilon = 1/100$ & \# updates & CPU time (s) \\ [0.5ex] 
\hline\hline
CRN & 1,000 & 2682 $\pm$ 1192 & 5950 $\pm$ 19123 & $\approx 1.26 \times 10^7$ & 46 \\ 
\hline
CRP & 1,000 & 2758 $\pm$ 569 & -2630 $\pm$ 9268 & $\approx 1.27 \times 10^7$ & 70 \\ 
\hline
CFD2 & 1,000 & 2655 $\pm$ 129 & 2640$\pm$ 1001 & $\approx 4.68 \times 10^6$ & 48 \\  
\hline\hline
CRN & 10,000 & 2453 $\pm$ 369 & 1505 $\pm$ 6120& $\approx 1.27 \times 10^8$ & 457 \\ 
\hline
CRP & 10,000 & 2783 $\pm$ 179 & 2627 $\pm$ 2937 & $\approx 1.27 \times 10^8$ & 672 \\ 
\hline
CFD2 & 10,000 & 2601 $\pm$ 40 & 2352 $\pm$ 282 & $\approx 4.68 \times 10^7$ & 483 \\ 
\hline\hline
CRN & 40,000 & 2386 $\pm$ 188 & 1069 $\pm$ 2984 & $\approx 5.07 \times 10^8$ & 1829 \\ 
\hline
CRP & 40,000 & 2745 $\pm$ 89 & 3593 $\pm$ 1468 & $\approx 5.07 \times 10^8$ & 2739 \\ 
\hline
CFD2 & 40,000 & 2582 $\pm$ 20 & 2512 $\pm$ 147 & $\approx 1.87 \times 10^8$ & 1931 \\ 
\hline
\end{tabular}
\caption{95\% confidence intervals for each of the finite difference methods (b), (c), and (d) for the computation in the mRNA and protein model of subsection \ref{mrna}. Note that the bias of the second finite difference can be seen when $\epsilon = 1/20$ (the actual value is 2496).  Also note that, though for a fixed number of estimates the CFD2 method is not the fastest method, it achieves a much smaller confidence interval. The number of updates and computational time for a fixed number of estimates are essentially independent of $\epsilon$ and so the reported values, here and throughout, are the average of the values for the two choices of $\epsilon$.}
\label{table:mp1} 
\end{table*}

\begin{table}
\centering 
\begin{tabular}{|c|c|c|c|} 
\hline 
Estimates & Approximation & \# updates & CPU time (s) \\ [0.5ex] 
\hline\hline
 1,000 & 2150 $\pm$ 2258 & $\approx 4.20 \times 10^6$  & 14\\ 
\hline
 10,000 & 2429 $\pm$ 729 & $\approx 4.19 \times 10^7$ & 135\\ 
\hline
 40,000 & 2176 $\pm$ 404 & $\approx 1.68 \times 10^8$  & 540\\
\hline
\end{tabular}
\caption{95\% confidence intervals for the LR method (d) for the computation in the mRNA transcription model computation of subsection \ref{mrna}. Even though this method is fastest per estimate, note that the variance (and so the width of the confidence interval) is large.}
\label{table:mp2} 
\end{table}

Perhaps more illustrative is Table \ref{table:mp3}, which compares the numbers of estimates and system updates as well as the time required to achieve a 95\% confidence interval of a set width.  These data give a good idea of the efficiency of the methods, as often one desires the estimate within a given tolerance.  We can see that the double coupled method is approximately 25 times faster than CRP, 73 times faster than the often used CRN method, over 100 times faster than the LR method, and over 125 times faster than IRN.  
 Note also that the double coupled method requires drastically fewer estimates to achieve the same confidence, so that, even though the computation of one double coupled estimate requires more time than most of the other methods, as can be seen in Table \ref{table:mp1}, the lower variance leads to very large time savings.

\begin{table}
\centering 
\begin{tabular}{|c|c|c|c|c|} 
\hline 
Method & Estimates & Approximation & \# updates & CPU time (s) \\ [0.5ex] 
\hline\hline
LR & 495,000 & 2506 $\pm$ 120&  $\approx 2.1 \times 10^9$ & 6619 \\ 
\hline
IRN & 190,000 & 2617 $\pm$ 120 & $\approx 2.4 \times 10^9$ & 7657 \\ 
\hline
CRN & 98,100 & 2572 $\pm$ 120 & $\approx 2.6 \times 10^8$ & 4489 \\ 
\hline
CRP & 22,200 & 2532 $\pm$ 120 & $\approx 2.8 \times 10^8$ & 1533 \\ 
\hline
CFD2 & 1150 & 2565 $\pm$ 120& $\approx 5.8 \times 10^6 $ &  61 \\ 
\hline
\end{tabular}
\caption{Required estimates, updates, and computational time needed for 95\% confidence intervals of $\pm$ 120 for all five methods on the computation of the mRNA transcription model computation of subsection \ref{mrna}. An $\epsilon$ of 1/20 was used for the finite difference methods.}
\label{table:mp3} 
\end{table}

\begin{figure}
\includegraphics[trim = .2in 2.5in .5in 2.3in, clip, width=2.8in]{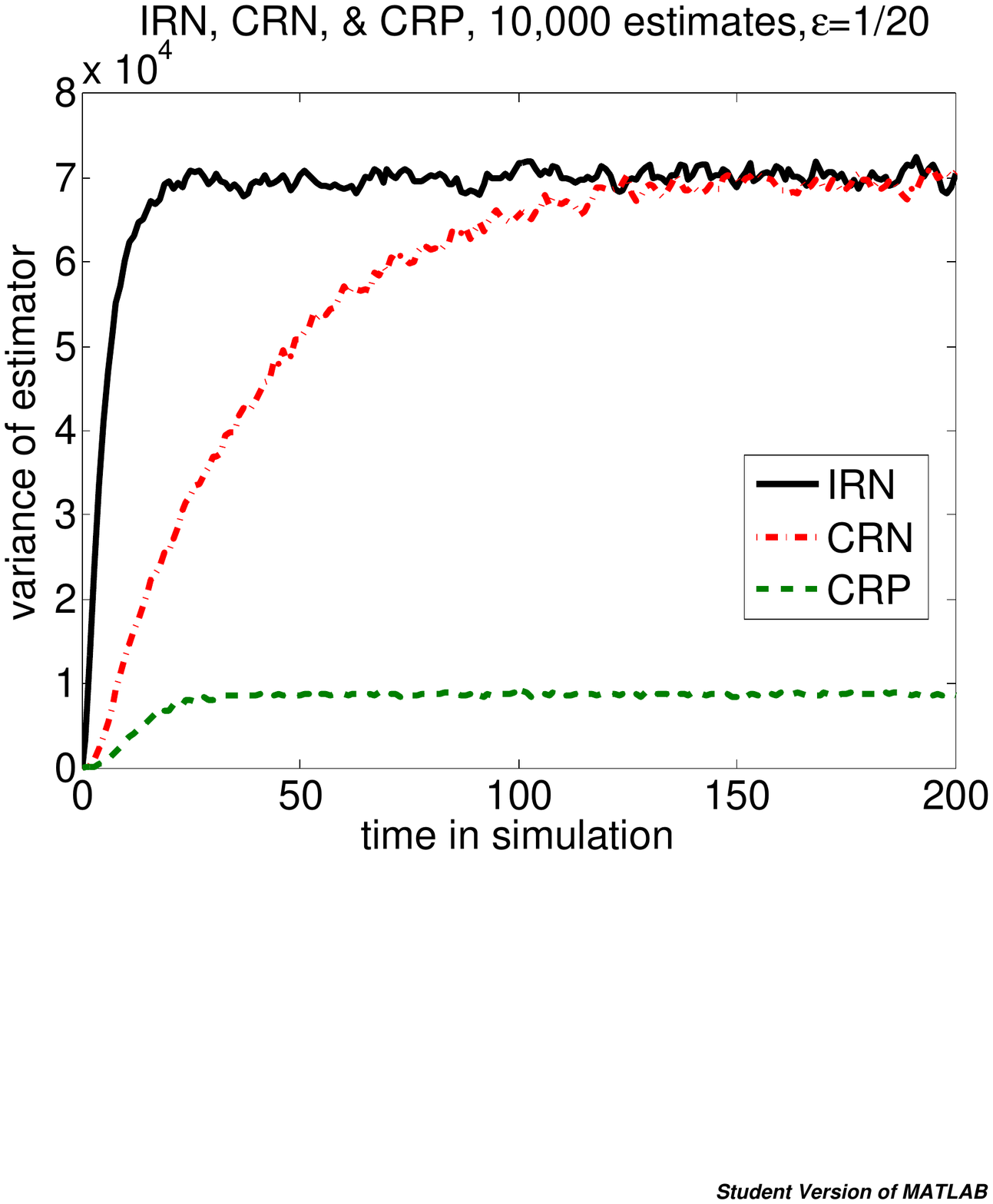}
\includegraphics[trim = .4in 2.5in .5in 2.4in, clip, width=2.6in]{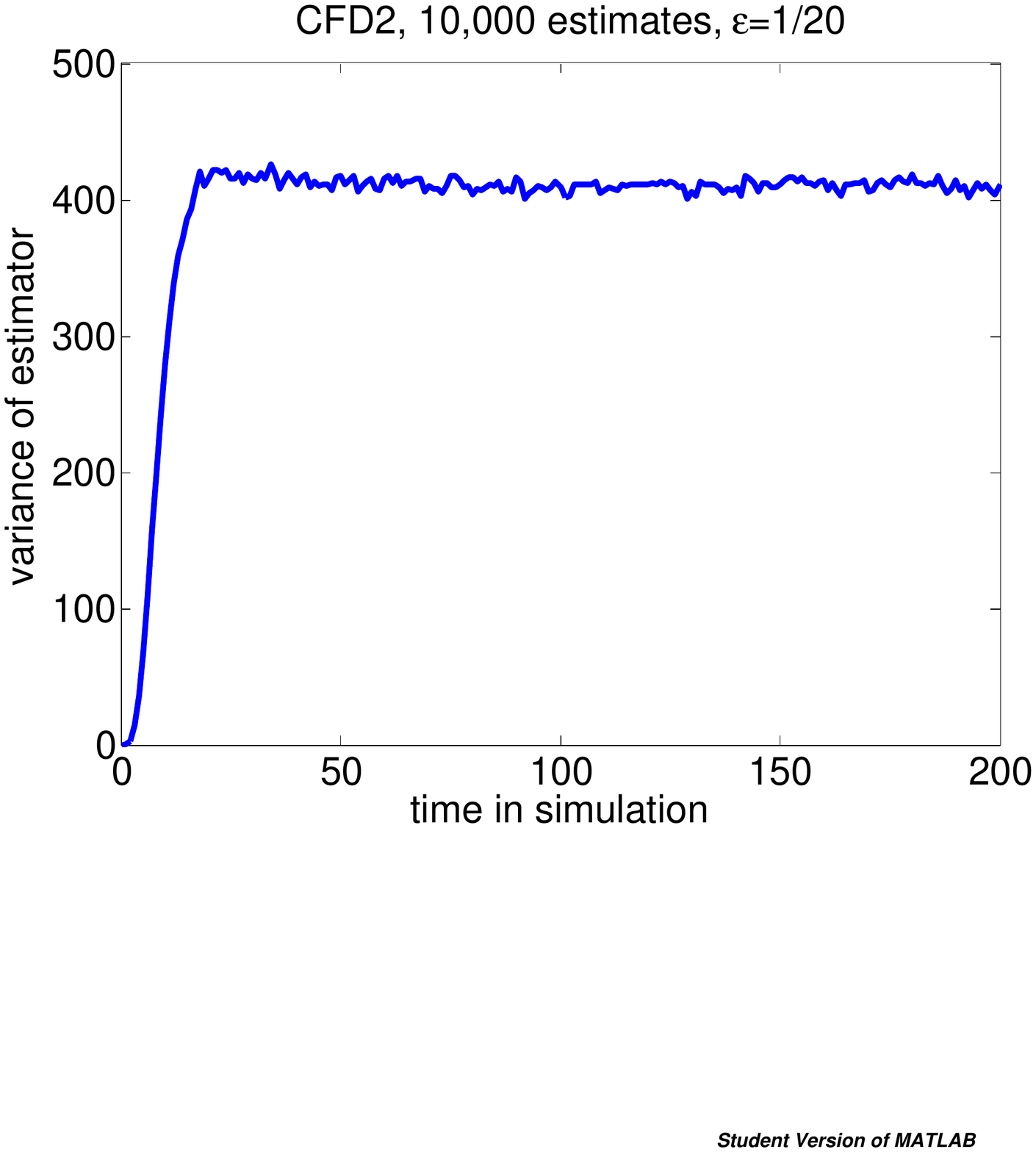}
\includegraphics[trim = .5in 2.5in .5in 2.3in, clip, width=2.6in]{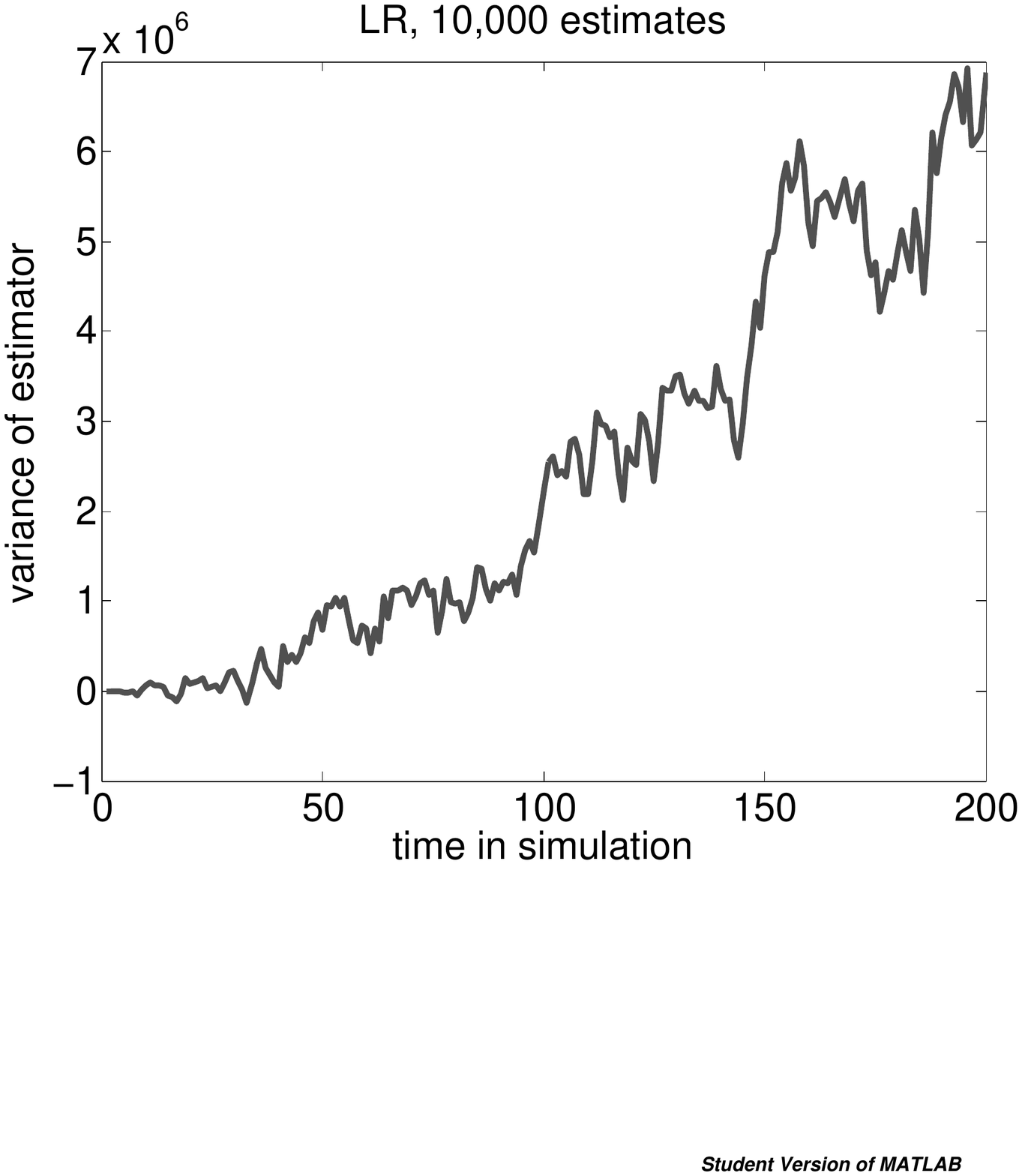}

\caption{Variance versus time of the estimators of the five different methods, applied to the calculation of $\frac{\partial^2}{\partial\theta^2}\mathbb{E}X_P(\theta,t)$ in the mRNA transcription model of subsection \ref{mrna}. Note that the scales are vastly different.}
\label{fig:mpplots}
\end{figure}

Finally, in Figure \ref{fig:mpplots} we  include a plot of the variance of the different estimators versus time.   Note that the scales on the plots are very different.  The plots corresponding to finite difference methods all appear to converge; the limiting value for the double coupled method, however, is over 20 times smaller than the CRP method, and over 170 times smaller than the CRN and IRN methods.  Note also that, as time increases, the CRN variance tends to the same value as the IRN method;  this is expected, since we expect the processes to decouple.  
The variance for CRP behaves similarly, converging to a number of approximately the same order of magnitude as the IRN method, though the value itself is significantly lower in this four-reaction model.  The plot for the LR method scales quadratically, as is expected by the form of the estimator (see Chapter VII.3 of \onlinecite{GlynnAsmussen2007}).
This shows that,  for moderate and large times, the double coupled method quickly becomes much more efficient then the other estimators.

\subsubsection{Mixed partial of protein abundance}\label{mixed}
We compare the five methods in the estimation of $\frac{\partial^2}{\partial\gamma\partial\theta}\mathbb{E}X_P(30)$ at $\theta= 1/4$ and $\gamma= 10$, which can be calculated exactly to be -31.8.  Table \ref{table:mixedmp} shows the approximations and computational complexity of these methods using $5,000$ estimates.

\begin{table}
\begin{tabular}{|c|c|c|c|c|} 
\hline 
Method & Estimates & Approximation & \# updates & CPU time (s) \\ [0.5ex] 
\hline\hline
IRN & 5,000 & 607 $\pm$ 923 & $\approx 8.43 \times 10^7$ & 264 \\ 
\hline
CRN & 5,000 & 191.5 $\pm$ 330 & $\approx 8.42 \times 10^7$ & 273 \\ 
\hline
CRP & 5,000 & 21.0 $\pm$ 96 & $\approx 8.41 \times 10^7$ & 365 \\ 
\hline
CFD2 & 5,000 & -33.8 $\pm$ 4& $\approx 2.25 \times 10^7$ & 238  \\ 
\hline
LR & 5,000 & -15.4 $\pm$ 113 & $\approx 2.10 \times 10^7$ & 73 \\ 
\hline\hline
LR & 17,000 & -61.8 $\pm$ 68 & $\approx 6.72 \times 10^7$ & 234 \\ 
\hline
\end{tabular}
\caption{95\% confidence intervals and computational complexity for all five methods, after $5,000$ estimates, for the computation of the mixed partial derivative in the mRNA transcription model as in subsection \ref{mixed}. An $\epsilon$ of 1/25 was used for the finite difference methods. Additionally, results from the LR method with CPU time approximately that of CFD2 are included for comparison. Actual value: -31.8.}
\label{table:mixedmp} 
\end{table}

\begin{figure}
\includegraphics[trim = .5in 2.5in .5in 2.4in, clip, width=3in]{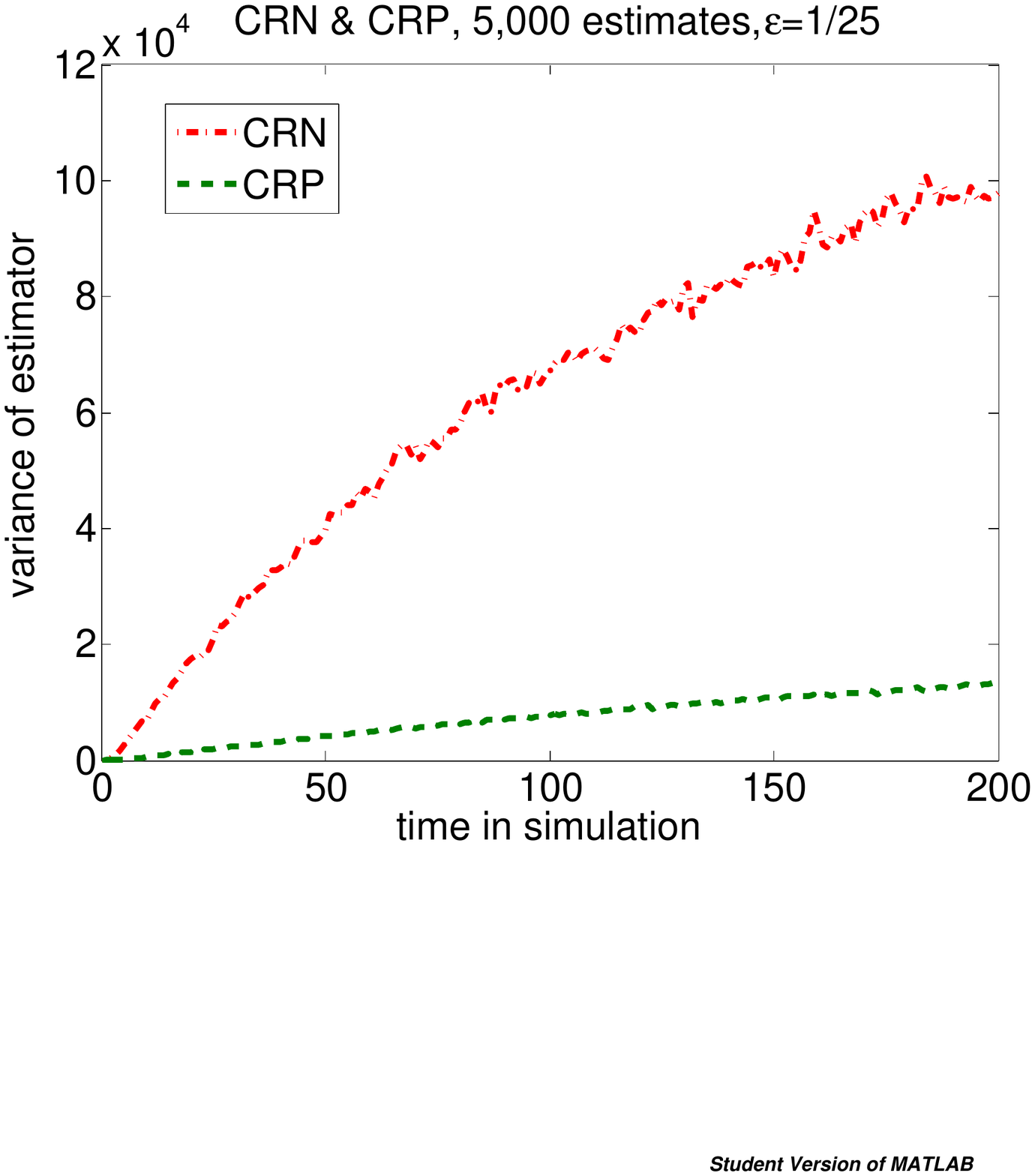}
\includegraphics[trim = .5in 2.5in .5in 2.2in, clip, width=3in]{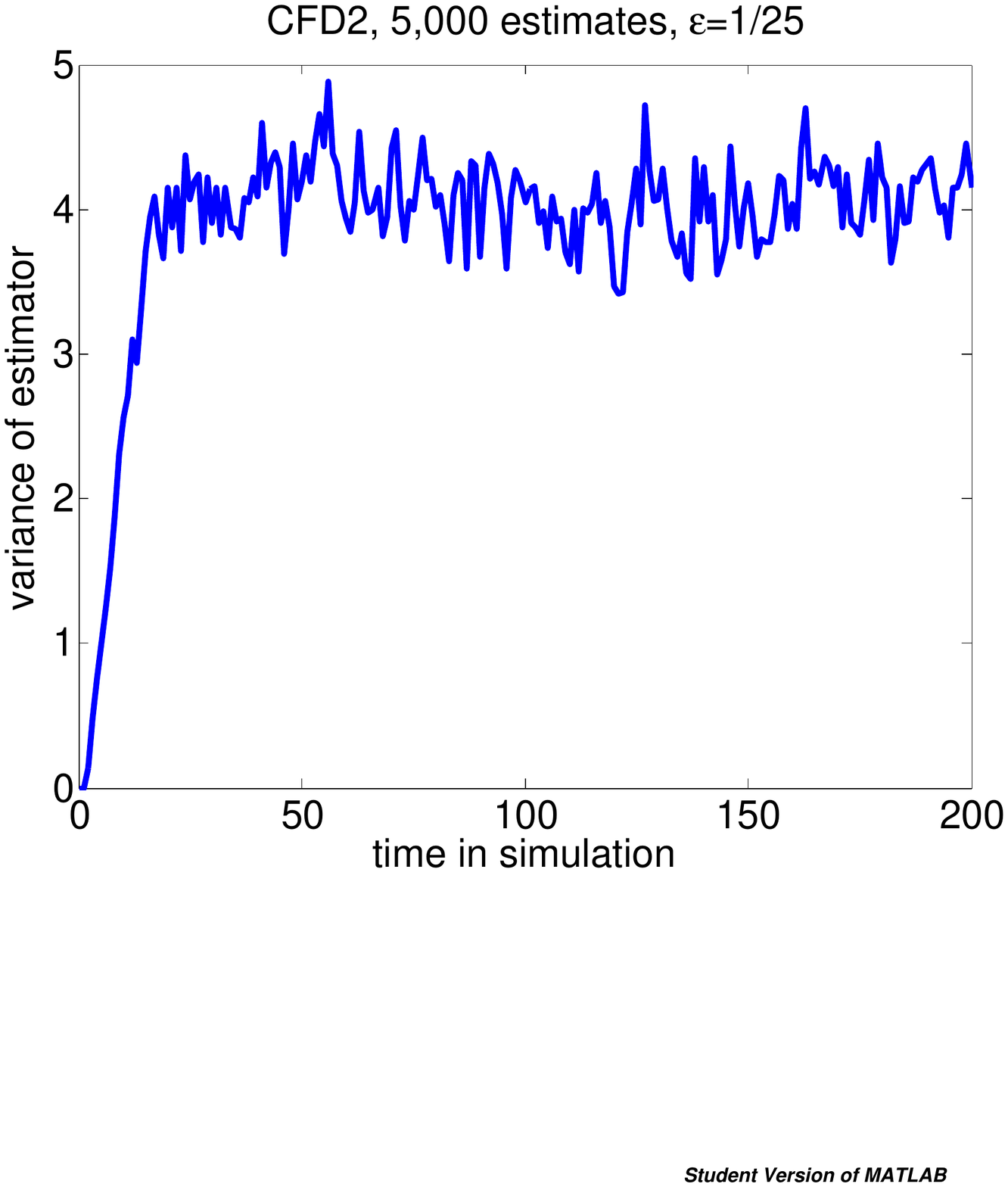}
\caption{Plots of variance over time for the CRN and CRP methods, and the CFD2 method, in computing the mixed partial derivative of the mRNA transcription model of subsection \ref{mixed}.  Note the very different scales.  For comparison, the IRN method plateaus at a variance of approximately $5\times 10^6$.}
\label{fig:mixed}
\end{figure}

Note that in this example, the LR method outperforms all methods, except CFD2, with respect to computation time.  Thus, for comparison, we have also included the results of a test using the LR method in which the CPU time is approximately the same as CFD2; note that the confidence interval for the CFD2 method is much smaller. Figure \ref{fig:mixed} shows variance plots of the CRN and CRP, and CFD2 methods over time in simulation.

\subsubsection{$2^{nd}$ derivative of the square of protein abundance with respect to $\theta$}\label{genf}
We also calculate, from the mRNA transcription model of Example 4.1, $\frac{\partial^2}{\partial\theta^2}\mathbb{E}(X_P(t)^2)$ at $t=5$ and $\theta= \frac{1}{4}$, with $\gamma=10$ and $X_0=0$.  Note here we are considering a function $f$ of the state space which is non-linear.

\begin{figure}
\includegraphics[trim = .5in 2.5in .5in 2.5in, clip, width=3in]{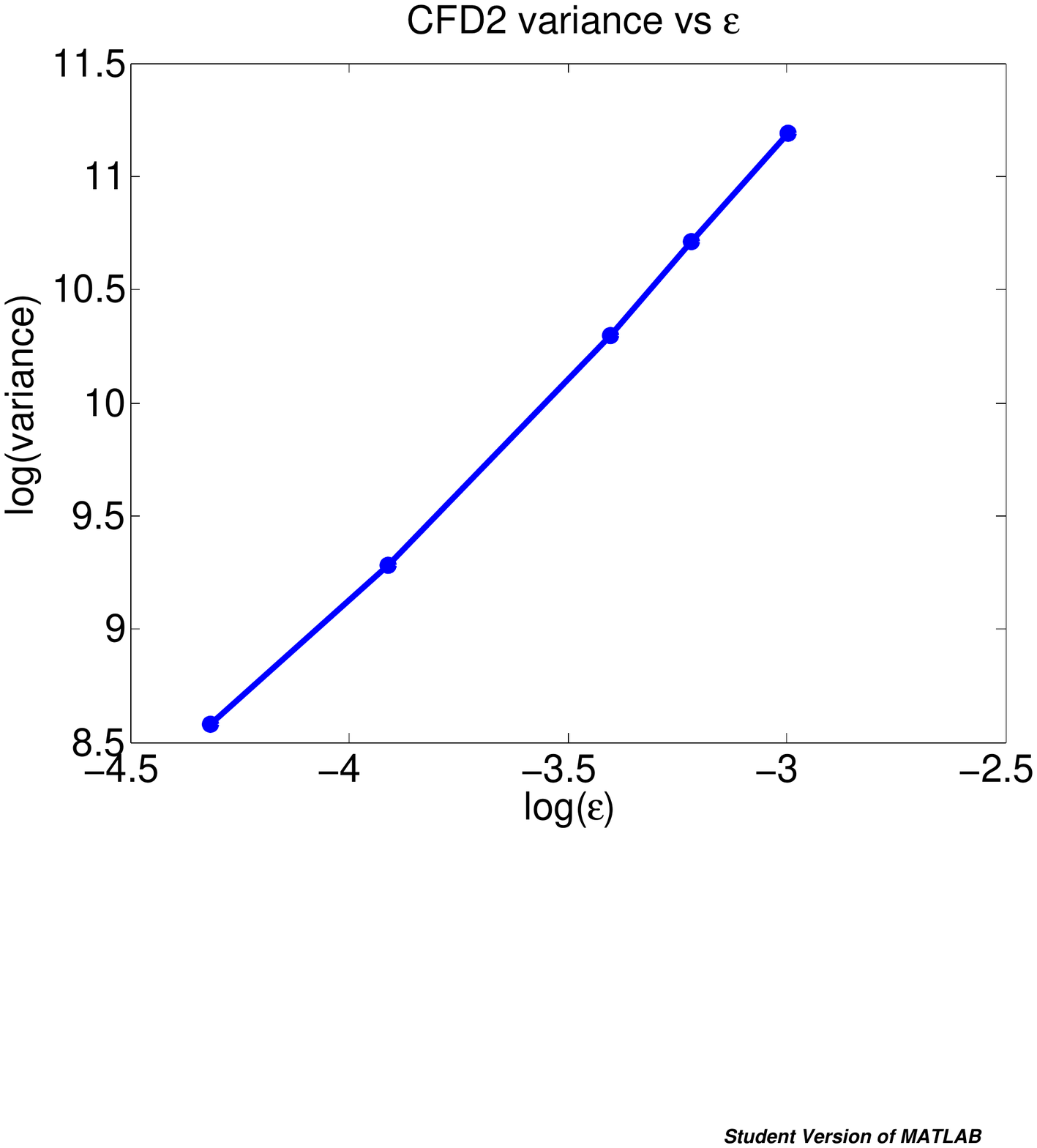}
\caption{This log-log plot of variance versus epsilon (100,000 estimates) for the mRNA transcription model computation of subsection \ref{genf} suggests that the CFD2 method gives an estimator of $O(\epsilon^{-2})$ even though the function $f$ of the system state is non-linear: the slope of the best fit line is 1.98.}
\label{fig:genf}
\end{figure}

In Figure \ref{fig:genf}, 
we plot the log of the variance of the numerator of the estimator (\ref{cdiff}) versus the log of epsilon.  Since we expect, for the double coupled CFD2 method, that this variance $V(\epsilon)$ should scale like $C\epsilon^p$ for some constants $C$ and $p$, we see that the slope of $\log(V(\epsilon))=\log(C)+p\log(\epsilon)$ from our simulations will suggest the value of $p$.  This plot suggests that $p=2$;
since the numerator of the estimator is then divided by $\epsilon^2$ in $d(\epsilon)$, this suggests a final variance of $O(R^{-1}\epsilon^{-2})$ for the estimator \eqref{eq:estimator} as discussed in Section \ref{model}.

For comparison, the slope of this log-log plot for the IRN method is zero, as the variance of the numerator does not depend on epsilon, giving a final variance of $O(R^{-1}\epsilon^{-4})$.  The slopes for the associated log-log plots
for the CRN and CRP estimators will 
vary with time (discussed further in subsection \ref{toggle}).

\begin{figure}
\includegraphics[trim = .5in 2.5in .5in 2.5in, clip, width=3in]{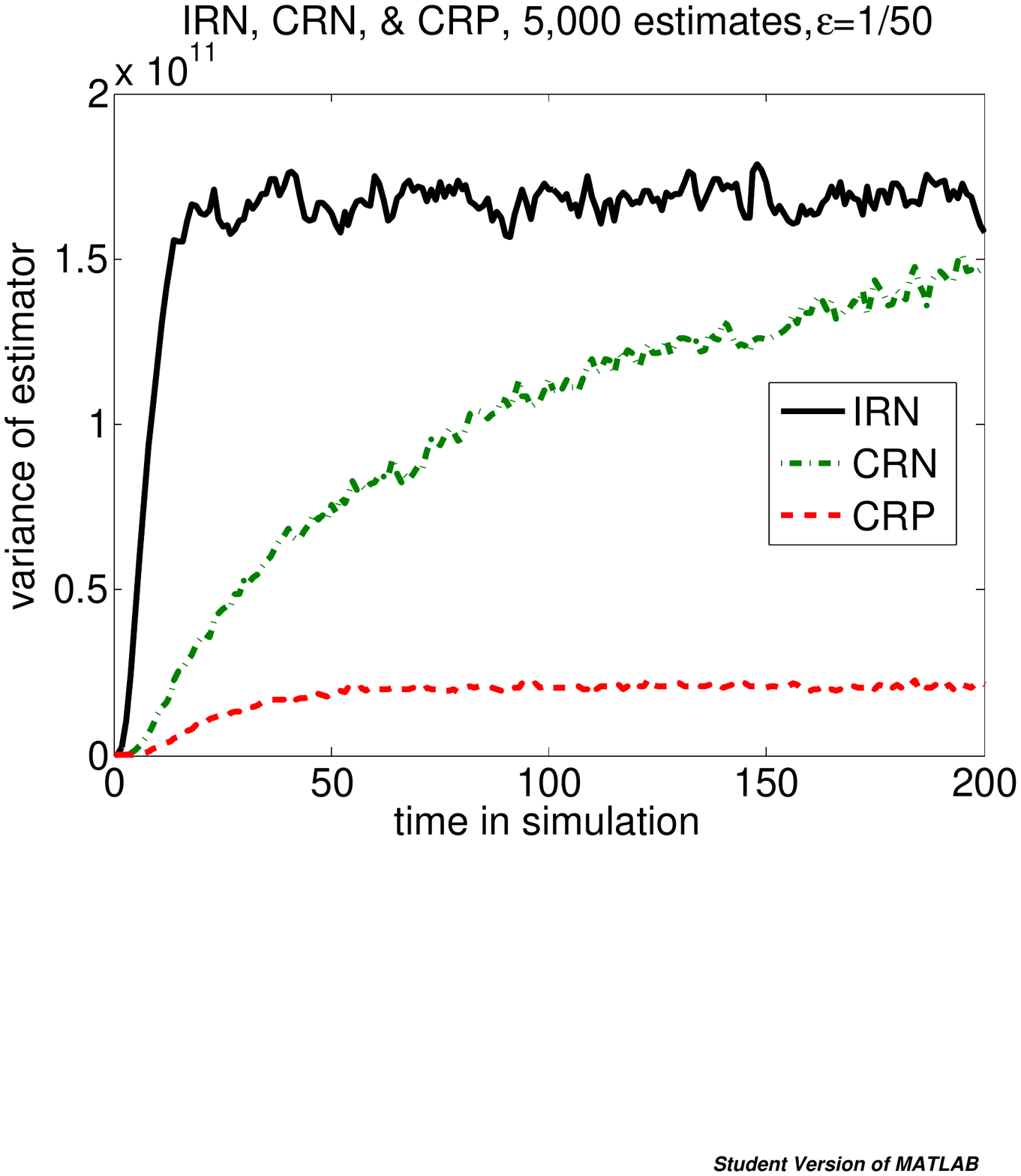}
\caption{Plot of variance over time for 5,000 estimates of the IRN, CRN, and CRP methods for the mRNA transcription model computation of subsection \ref{genf}. The variance of the CFD2 method is too small to be seen at this scale; at time 200 it is approximately $3.5\times 10^8$. }
\label{genf2}
\end{figure}

The general behavior of the variances over time for the IRN, CRN, and CRP methods can be seen in Figure \ref{genf2}.

\subsection{Quadratic Decay}\label{decay}
\begin{figure}
\includegraphics[trim = .5in 2.4in .5in 2.2in, clip, width=3in]{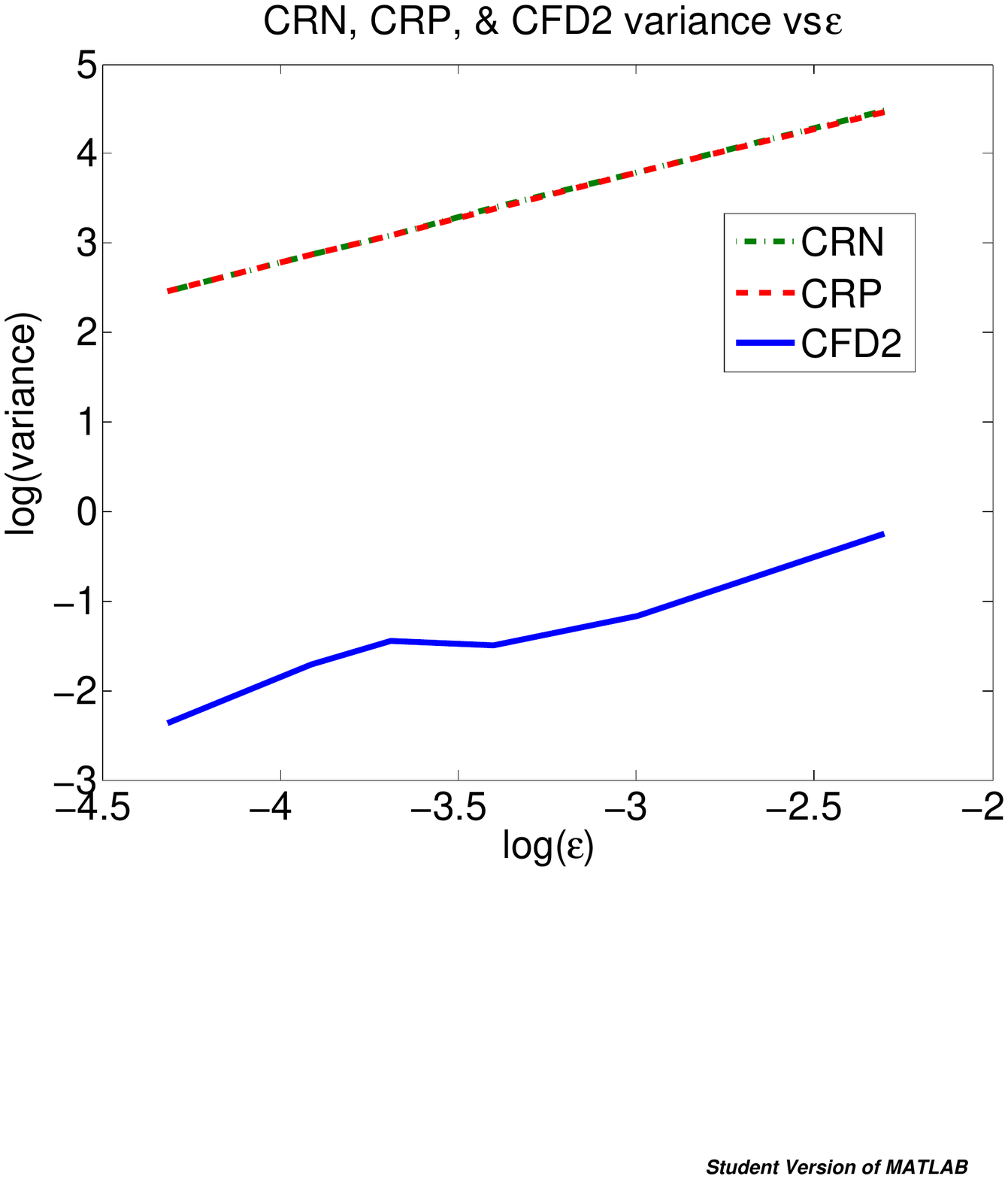}
\caption{This log plot of variance versus epsilon (each point computed to the first of 300,000 estimates or a confidence of $\pm10$) for the decay model of subsection \ref{decay} suggests that the CFD2 method gives an estimator of only $O(\epsilon^{-3})$: the slope of the best fit line is approximately 0.97.  While the CRN and CRP methods also give an estimator of this same rate (the slope of the best fit lines are both $\approx$.99, and in fact the lines are on top of each other), the variance of the estimates from CRN and CRP are significantly higher than those from the CFD2 method, as can be seen by the wide gap between the above curves. }
\label{fig:decaylog}
\end{figure}

In order to demonstrate that the $O(\epsilon^2)$ convergence rate seen in the previous examples does not universally hold, we consider a pure decay process of a population $X_t$, so that the sole reaction has $\zeta=-1$ and quadratic propensity $\lambda(\theta,X_t(\theta))=\theta X_t(\theta)(X_t(\theta)-1)$, and calculate $\frac{\partial^2\E X_t(\theta)}{\partial\theta^2}$ with $\theta=1$ and with initial population $X_0(\theta)=2000$. Figure \ref{fig:decaylog} gives a log-log plot of variance versus epsilon at time $0.001$. Since it suggests $p=1$, this demonstrates that, in this case, the double coupled method provides only $O(R^{-1}\epsilon^{-3})$ convergence as discussed in Section \ref{sec:model}, showing that rate to be sharp.

\begin{figure}
\includegraphics[trim = .2in 2.4in .5in 2.2in, clip, width=3in]{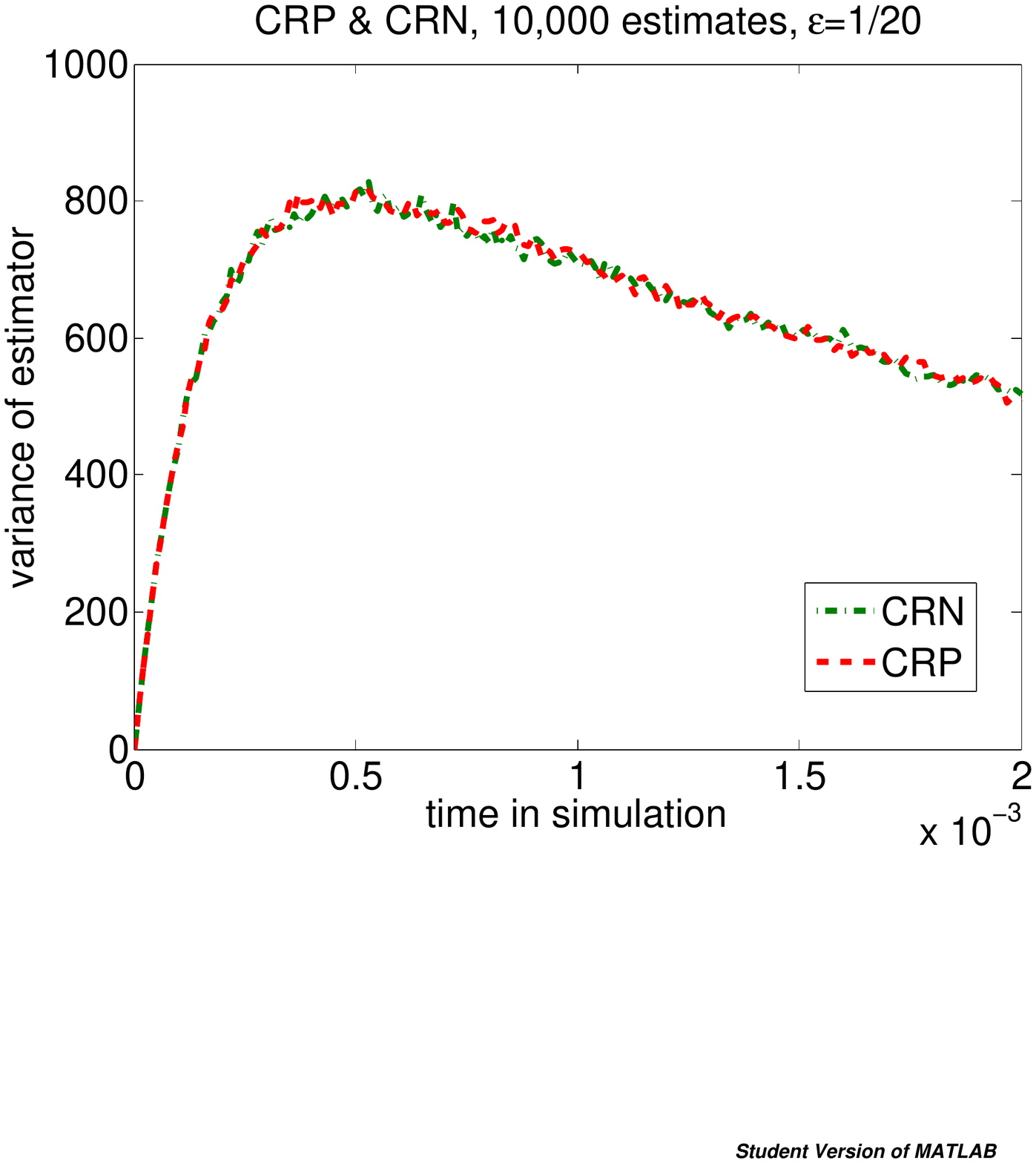}
\includegraphics[trim = .2in 2.4in .5in 2.2in, clip, width=3in]{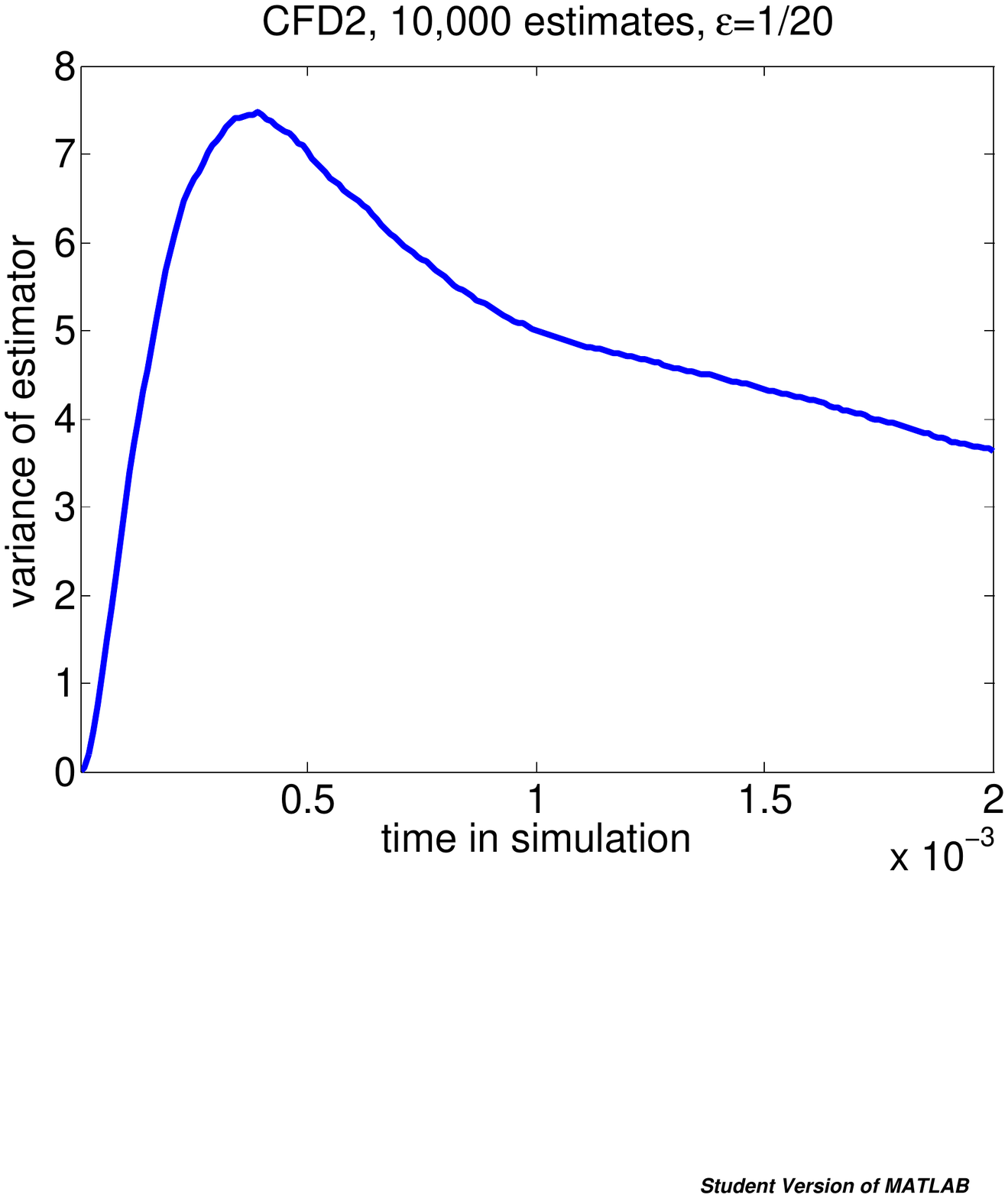}
\caption{The behavior over time of the variance of the estimates of the CRN, CRP, and CFD2 methods on the quadratic decay model of subsection \ref{decay}.  Note that the variance for the CFD2 method is 100 times smaller than the other two methods, which, as expected, act the same on this model.  An $\epsilon$ of 1/20 was used and 10,000 estimates were run. The plot of the IRN variance is similar in shape but with a peak variance of 3.1$\times 10^4$.}
\label{fig:decayvar}
\end{figure}

As demonstrated in Table \ref{table:decay} and in Figure \ref{fig:decayvar}, however, the double coupled method is still significantly more efficient than existing methods on this model.

\begin{table*}
\centering 
\begin{tabular}{|c|c|c|c|c|c|} 
\hline 
Method & $\epsilon$ & Estimates & Approximation & \# updates & CPU time (s) \\ [0.5ex] 
\hline\hline
LR & n/a & 10,000 & 1240 $\pm$ 1070 & $\approx 1.3 \times 10^7 $ & 9 \\
\hline
IRN &1/20& 10,000 & 555 $\pm$ 218 & $\approx 4.8 \times 10^7 $ & 35 \\ 
\hline
CRN &1/20& 10,000 & 585 $\pm$ 52 & $\approx  4.0 \times 10^7$ & 30 \\ 
\hline
CRP &1/20& 10,000 & 584 $\pm$ 52 & $\approx  4.0 \times 10^7$ & 30 \\ 
\hline
CFD2 &1/20& 10,000 & 592 $\pm$ 5& $\approx 1.4\times 10^7$ & 90  \\ 
\hline\hline
CRN &1/20& 272,000 & 588 $\pm$ 10 & $\approx  1.1 \times 10^9$ & 813 \\ 
\hline
CRP &1/20& 271,000 & 589 $\pm$ 10 & $\approx  1.1 \times 10^9$ & 862 \\ 
\hline
CFD2 &1/20& 1,950 & 592 $\pm$ 10& $\approx 2.7 \times 10^6$ & 17 \\ 
\hline\hline
CRN &1/50& 169,500 & 543 $\pm$ 50 & $\approx  6.8 \times 10^8$ & 511 \\ 
\hline
CRP &1/50& 169,000 & 515 $\pm$ 50 & $\approx  6.8 \times 10^8$ & 510 \\ 
\hline
CFD2 &1/50& 1,800 & 605 $\pm$ 50 & $\approx  2.5 \times 10^6$ & 16 \\ 
\hline
\end{tabular}
\caption{Estimates, $\epsilon$ used, and updates and computational time needed for the given 95\% confidence intervals for all methods  for $\frac{\partial^2\E X_t(\theta)}{\partial\theta^2}$ at $t=0.001$ for the quadratic decay model of subsection \ref{decay}. The upper half of the table shows the relevant results after the simulation of 10,000 estimates.  The lower half of the table shows the results of simulations run until the estimate had a confidence interval of a desired width.  The IRN and LR methods were unable to achieve these precisions due to memory constraints. Note again the equivalence of the CRN and CRP methods on a single reaction model.}
\label{table:decay} 
\end{table*}

\subsection{Genetic Toggle Switch}

Finally, we consider a model of a genetic toggle switch that also appeared in \onlinecite{Khammash2010} and \onlinecite{Anderson2011},
\[
\emptyset \underset{1}{\overset{\lambda_1}{\rightleftarrows}} A \hspace{.15cm}, \hspace{.5cm} \emptyset \underset{1}{\overset{\lambda_2}{\rightleftarrows}} B 
\]
where 
\[
	\lambda_1(t)=\frac{b}{1+X_B(t)^{\beta}}\quad \text{and} \quad \lambda_2(t)=\frac{a}{1+X_A(t)^{\alpha}},
	\]
	 and where $X_A(t)$ and $X_B(t)$ denote the number of gene products from two interacting genes.  Note that each gene product inhibits the growth of the other.  

We take parameter values of $b = 50, \beta= 2.5, a=16$ and will differentiate with respect to $\alpha$. Note that this model does \textit{not} follow mass action kinetics, or have linear propensities.  In subsection \ref{toggle}, we consider a second derivative of $\E X_B$ at a fixed time, while in subsection \ref{functional}, we consider a second derivative of the expected \textit{time average} of $X_A$ up to a given time, which is a functional of the path of $X_A$ rather than simply $X_A$ at some terminal time.

\subsubsection{$2^{nd}$ derivative of abundance of B with respect to $\alpha$}\label{toggle}

\begin{figure}
\includegraphics[trim = .5in 2.2in .5in 2.2in, clip, width=3in]{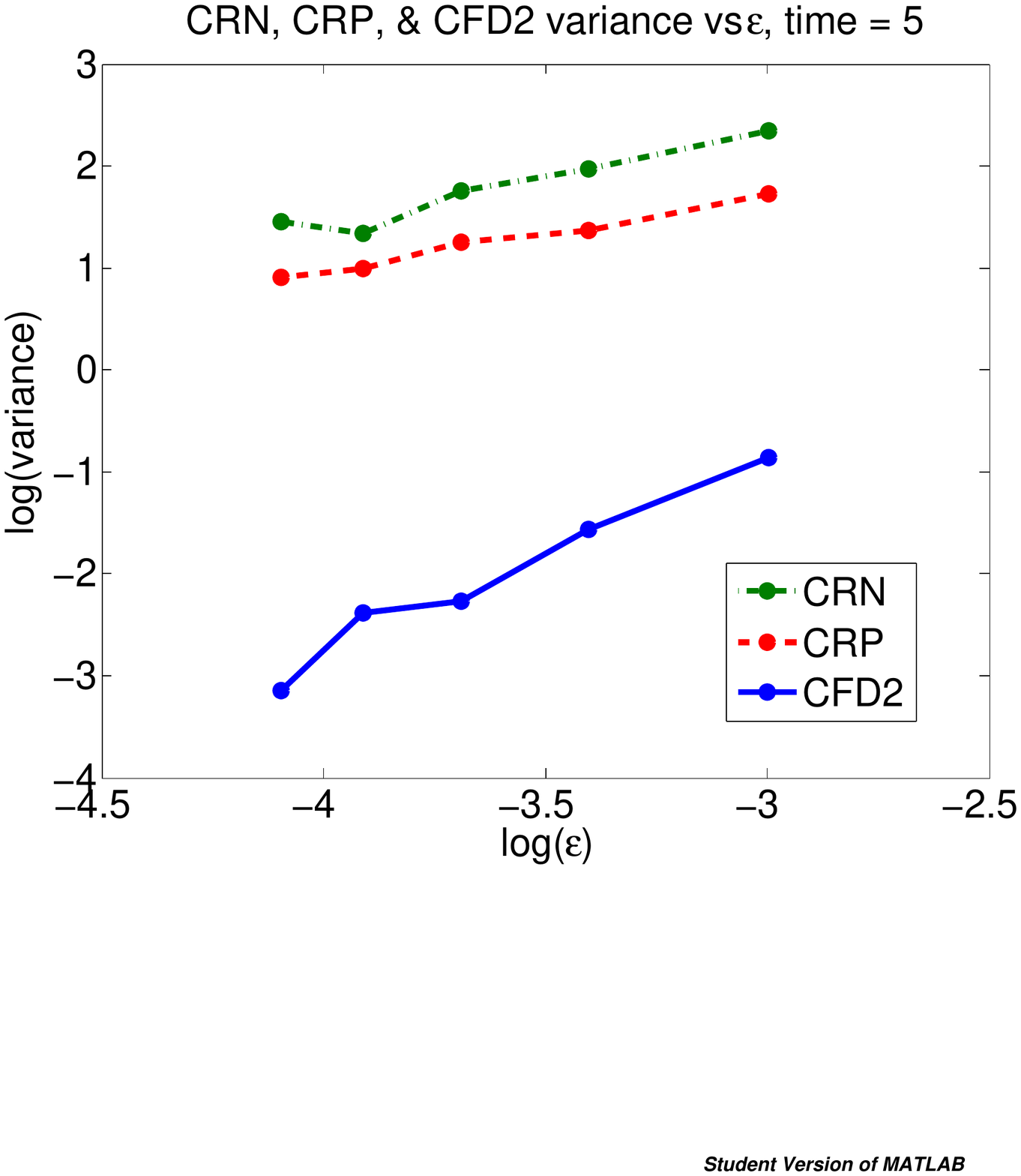}
\includegraphics[trim = .5in 2.4in .5in 2.2in, clip, width=3in]{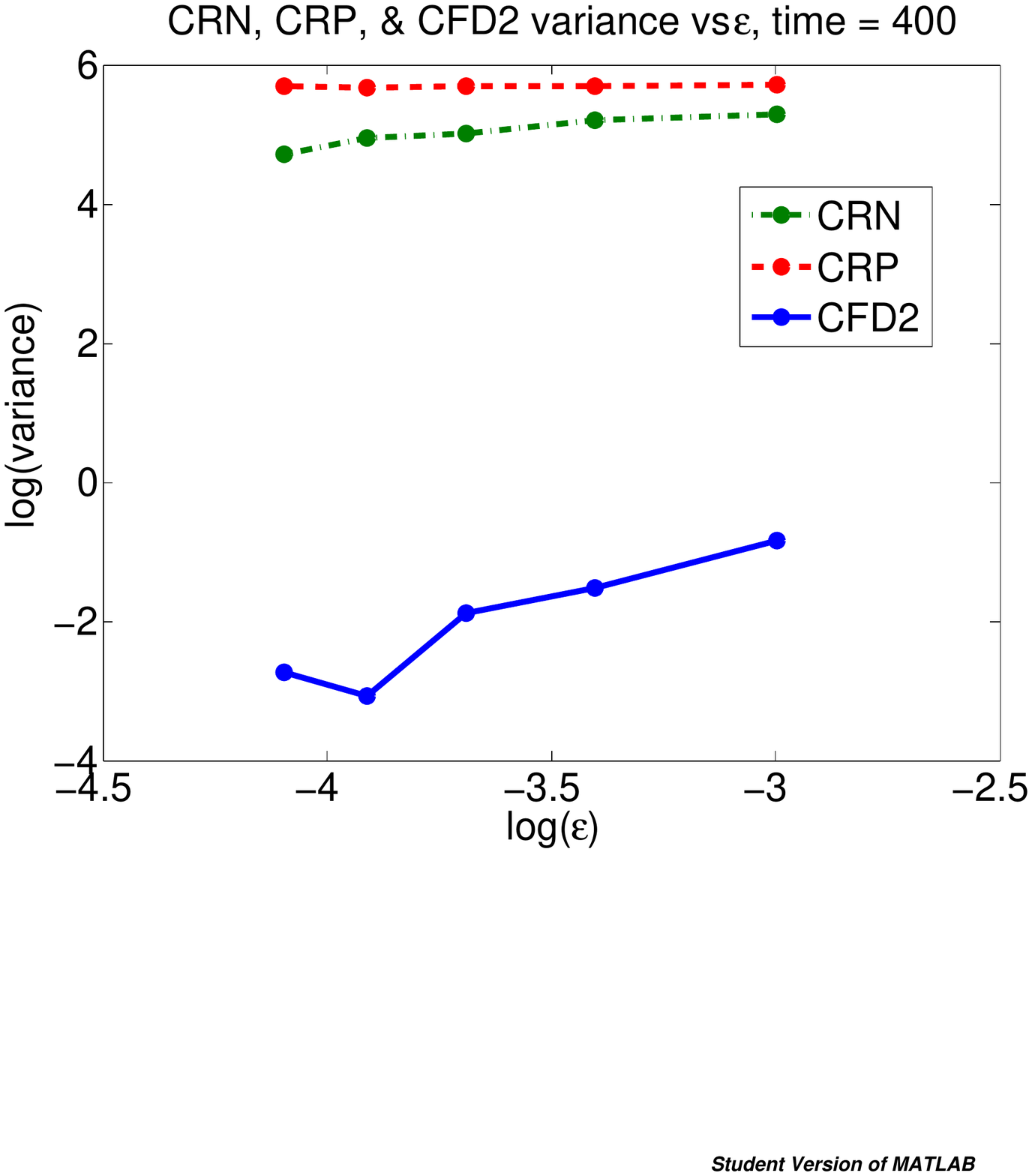}
\caption{These log-log plots (25,000 estimates) of variance versus epsilon for computation on the gene toggle model as in subsection \ref{toggle}, at two different times, suggest that the double coupled method gives an estimator of $O(\epsilon^{-2})$ even though two of the intensities are nonlinear: the slope of the best fit line for the CFD2 method is approximately 2 (=1.97) at both times.  The slope for the CRP and CRN methods, on the other hand, are approximately  .74 and .90 respectively at time 5, but are only around  .03 and .49 at time 400.}
\label{fig:toggle}
\end{figure}

\begin{figure}[hbt]
\includegraphics[trim = .4in 2.5in .5in 2.2in, clip, width=2.9in]{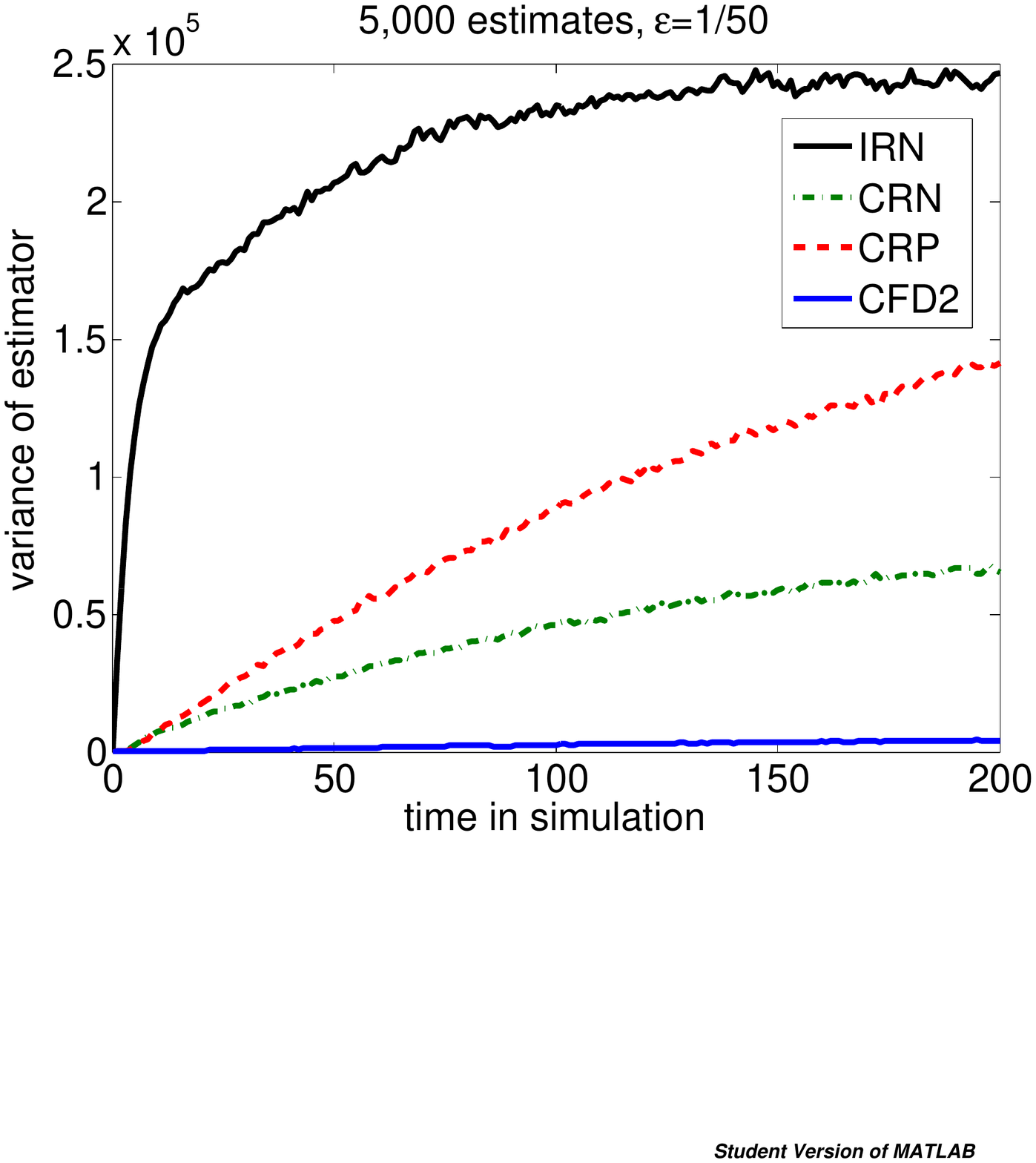}
\includegraphics[trim = .3in 2.5in .5in 2.2in, clip, width=3in]{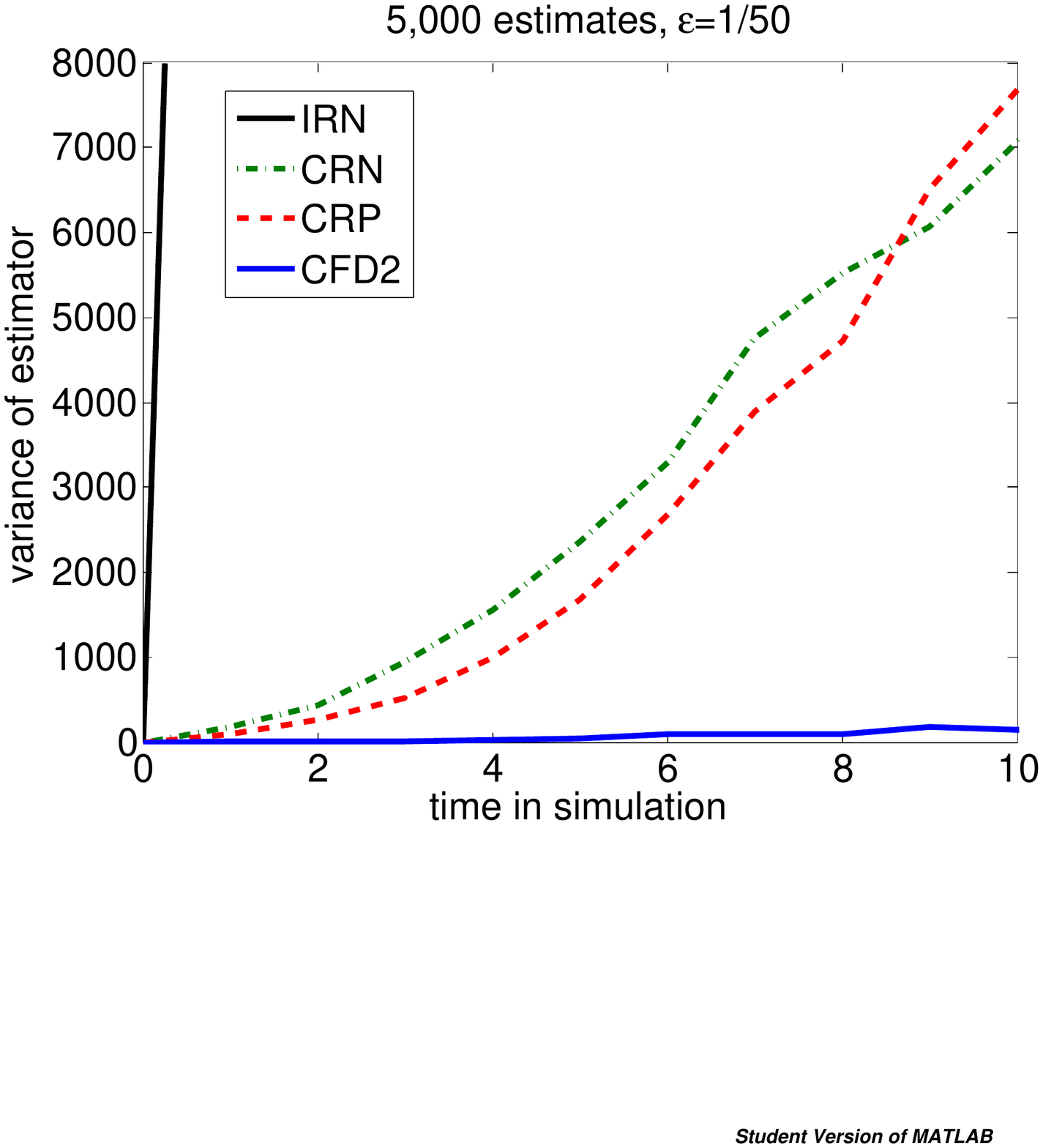}
\caption{Plots of variance over time for 10,000 estimates of the finite difference methods  for the gene toggle model as in subsection \ref{toggle}; the top includes time up to 200, while the bottom provides a close-up view of the plot for times less than 10. 
The value of the CFD2 variance at time 200 is approximately 4,000, while the CRN variance is approximately 65,000.}
\label{fig:toggle2}
\end{figure}

We estimate $\frac{\partial^2\mathbb{E}X_B(\alpha,t)}{\partial\alpha^2}$ at $\alpha= 1$ and at two times, 5 and 400. In Figure \ref{fig:toggle}, we plot the log of the variance of the numerator of the estimator (\ref{cdiff}), using CFD2, versus the log of the perturbation epsilon.  As in subsection \ref{mixed}, the plot clearly suggests that $p=2$.  
 We also plot the same quantity using CRP and CRN.  These slopes, on the other hand, vary with time.  For small times both slopes are close to one, but as time increases the slopes decrease, until, for very large times, they are close to zero.  This corresponds with the fact that for large times the variances of the CRP and CRN estimates converge to values on the order of the IRN estimate variance, which, as previously noted, is independent of the value of epsilon.
The general behavior of the variances over time can be seen in Figure \ref{fig:toggle2}, where it is seen that CFD2 has a variance that is 16 times lower than CRN and 36 times lower than CRP.  Further, we note that for this model the CRP method outperforms the CRN method for small times, while for larger times CRN outperforms CRP.

\subsubsection{$2^{nd}$ derivative of time average of abundance of $A$ with respect to $\alpha$}\label{functional}
\begin{figure}
\includegraphics[trim = .2in 2.5in .5in 2.2in, clip, width=3.25in]{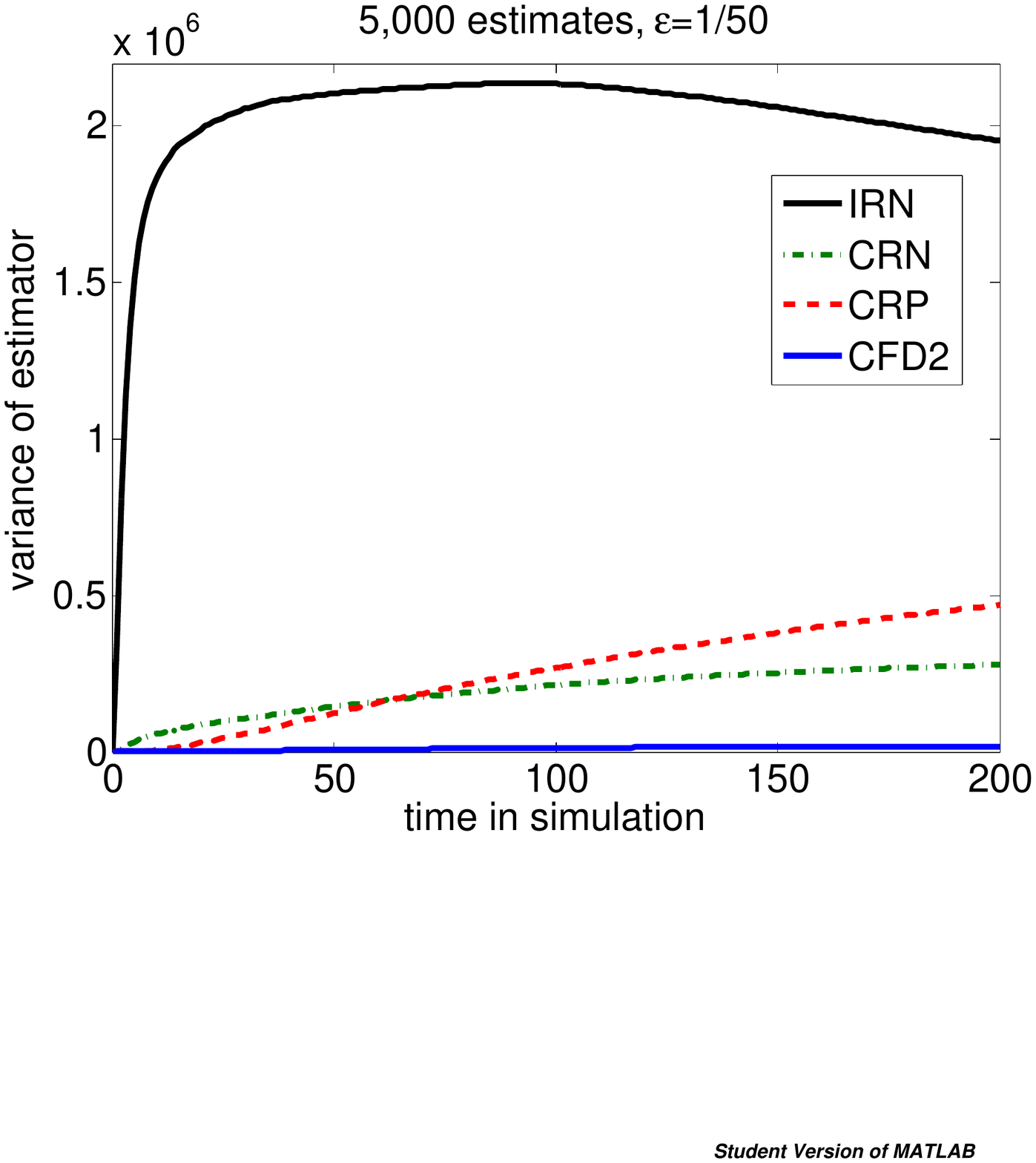}
\includegraphics[trim = .2in 2.5in .5in 2.4in, clip, width=3.25in]{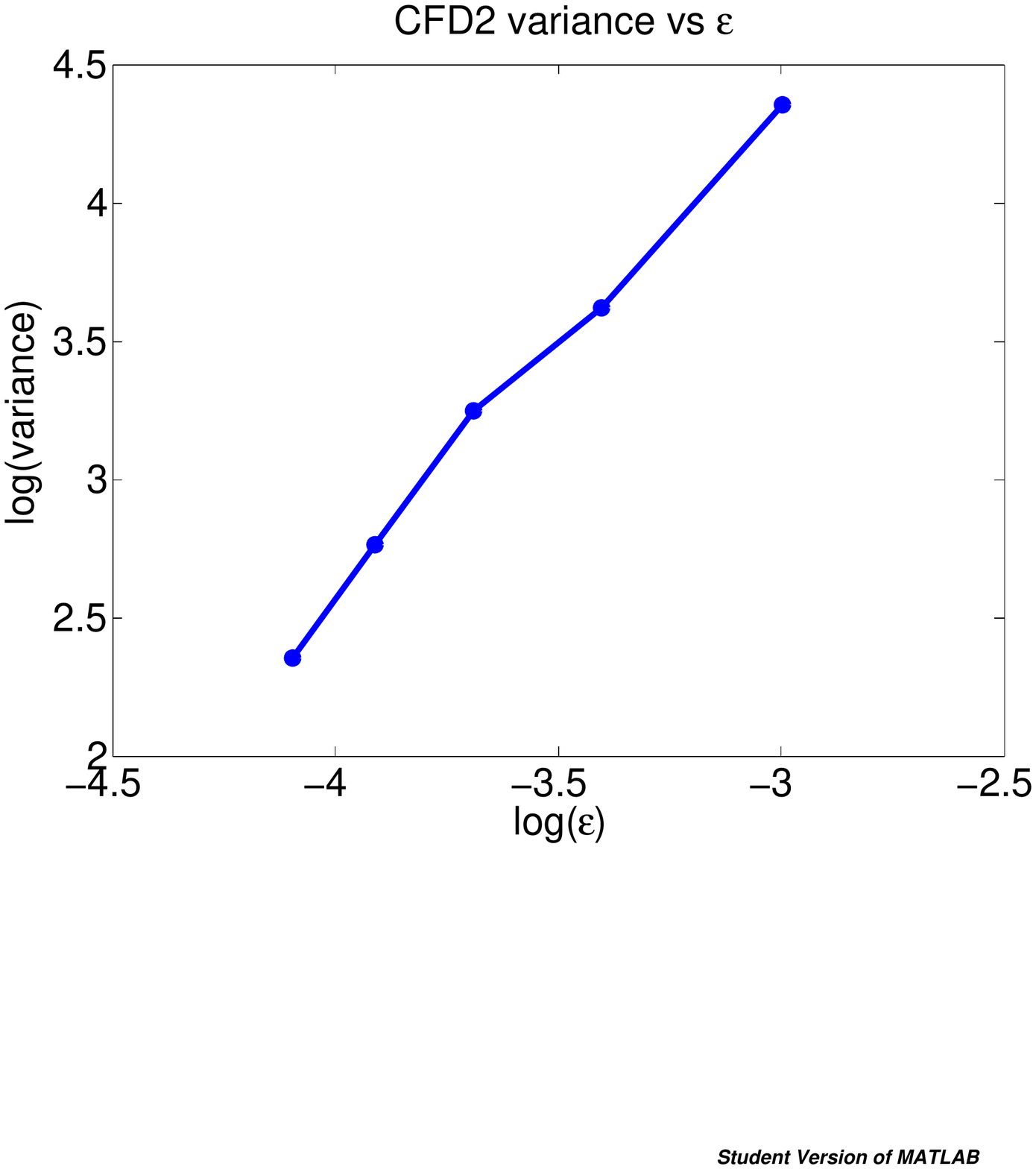}
\caption{At top, a plot of variance over time for 5,000 estimates for the finite difference methods for the path functional computation on the gene toggle model in subsection \ref{functional}. The value of the CFD2 variance at time 200 is approximately 19,000, while the CRN variance is approximately 282,000. At bottom, a log-log plot (2,000 estimates) of variance versus epsilon for this computation suggests that the double coupled method gives an estimator converging faster than $O(\epsilon^{-3})$ in this computation as well: the slope of the best fit line for the CFD2 method is approximately 1.78.}
\label{fig:funcvar}
\end{figure}

Finally, while this was not discussed in the paper, we include an example computing a sensitivity of a path functional.  That is, the quantity we wish to study is a function of the path of the process $X(s)$ for $s\leq t$, rather than just the terminal value $X(t)$.  The only difference in implementation is the need to compute this quantity during the simulation of the path (or to store the path for the computation after its simulation).  Table \ref{table:functional} shows the estimates of $\frac{\partial^2}{\partial \alpha^2}\E \, t^{-1}\int_0^t X_A(s) ds$ at $t=30$ using the various finite difference methods, demonstrating the advantage of the double coupled method for these path functional quantities as well. Additionally, Figure \ref{fig:funcvar} shows that the overall behavior of the variances of the three finite difference methods remains the same as in the previous examples.

\begin{table}
\centering 
\begin{tabular}{|c|c|c|c|c|} 
\hline 
Method & Estimates & Approximation & CPU time (s) \\ [0.5ex] 
\hline\hline
IRN & 100,000 & -13.8 $\pm$ 621 & 2240 \\ 
\hline
CRN & 100,000 & -274 $\pm$ 146 &  1441 \\ 
\hline
CRP & 100,000 & -215 $\pm$ 107 &  3035 \\ 
\hline
CFD2 & 100,000 &  -222 $\pm$ 26 & 2722  \\ 
\hline
\end{tabular}
\caption{95\% confidence intervals and computational complexity for each of the methods (a) through (d), after $100,000$ estimates, for the time average computation at $t=30$ on the gene toggle model of subsection \ref{functional}. An $\epsilon$ of 1/50 was used.}
\label{table:functional} 
\end{table}

\section{Conclusions and future work}
\label{sec:conclusion}
We have introduced a new, efficient method for the computation of second derivative sensitivities for discrete biochemical reaction networks.  Through several numerical examples we have demonstrated its advantage over existing methods, both in simple scenarios and in more realistic systems, including several examples in which the system contained nonlinear propensities, or in which the relevant quantity to be studied involved a nonlinear function $f$ or even a path functional of the system state.
Future work will include proving analytical bounds on the variance of the estimator given by the new method and exploring conditions in which a better convergence rate is achieved, as well as finding efficient algorithms to simultaneously compute all of the second order sensitivities of models with a large number of parameters.
Another avenue of future work will involve incorporating algorithms for the computation of second derivatives into the optimization methods discussed in the introduction in the context of parameter estimation.

\section*{Acknowledgements} 

Wolf was supported from NSF grants DMS-1009275 and DMS-1106424.  Anderson was supported from NSF grant DMS-1009275.  We thank James Rawlings for discussions that motivated this work.

\bibliography{doublecouple}

\end{document}